\documentclass[reprint, 
groupedaddress,
 amsmath,amssymb,
 aps, physrev,
]{revtex4-2}

\usepackage{comment}
\usepackage{subcaption}
\usepackage{graphicx}
\usepackage{dcolumn}
\usepackage{bm}

\begin{document}

\title{\textbf{Analytical Model of Clock Drift in Quantum Key Distribution and a Simple Synchronization Algorithm} 
}%

\author{L.~Millet$^{1,2}$, B.~Korzh$^{2}$, R.~Thew$^{2}$, and G.~Boso$^{1}$}
\email{gianluca.boso@idquantique.com}

\affiliation{$^{1}$ID Quantique SA, CH-1227 Geneva, Switzerland}
\affiliation{$^{2}$Department of Applied Physics, University of Geneva, CH-1205 Geneva, Switzerland}
\date{\today}

\begin{abstract}
Clock synchronization is critical for maintaining low error rates in quantum key distribution. Here, we describe how a frequency mismatch between the transmitter and receiver clocks affects the quantum bit error rate in quantum key distribution, and derive from this model a simple synchronization algorithm together with clock stability requirements for practical operation. Our algorithm continuously compensates for both frequency mismatch and time-offset fluctuations directly from detection timestamps. It does not require a dedicated synchronization channel or auxiliary qubit sequence, converges from a large frequency mismatch within approximately one second of photon acquisition, and remains effective in low-photon-count regimes (more than 30\,dB of channel loss) using standard hardware. We validate our approach by demonstrating successful key exchange over 100\,km of fiber and continuous operation over 24 hours in a 16\,km metropolitan network using commercial systems, with performance equivalent to using a service channel for clock synchronization. 
\end{abstract}

\maketitle

Quantum key distribution (QKD) enables information-theoretically secure communication through the principles of quantum mechanics, unlike classical cryptographic methods that rely on computational assumptions for their security\,\cite{gisin2002,Pirandola2020}. In a practical QKD system, various imperfections can degrade performance\,\cite{Diamanti2016,Xu2020}. In particular, a frequency mismatch between the local clocks of the transmitter (Alice) and receiver (Bob) can introduce errors in the secret key, while an unknown absolute time offset can prevent associating Bob's detections with Alice’s qubit sequence. To mitigate these effects, a reference clock signal is commonly shared between Alice and Bob over a dedicated channel\,\cite{Korzh2015}. However, the constraints associated with such a channel hinder seamless integration into existing telecommunication networks, and synchronization methods operating directly from the quantum channel are required for wider deployment.

Several approaches have been proposed to eliminate the dedicated clock channel in QKD\,\cite{Bourgoin2015,Takenaka2017SatelliteGround,Caldero2020,Agnesi2020,Wang2021,Shakhovoy2023,Zahidy2023,Spiess2024,Krause2025ClockOffset}. In these approaches, frequency recovery typically relies on fast Fourier transforms, frequency scanning, fitting, or multiplexing on the quantum channel, whereas time-offset recovery usually implies cross-correlation with a known qubit sequence. Current approaches can involve heavy computations and low jitter hardware, which limits compatibility with commercial systems. In this work, we first present a detailed analytical study of how a frequency mismatch between the local clocks of Alice and Bob alters the photon detection-time probability distribution. This analysis naturally leads to a simple and lightweight synchronization algorithm that estimates and compensates for the clock frequency mismatch directly from detection timestamps. By analytically relating the quantum bit error rate (QBER) to the frequency mismatch, we establish explicit requirements on clock stability for practical operation, which are shown to be compatible with low-cost oscillators. Based on these results, we implement the proposed algorithm using standard hardware and demonstrate its robustness in a deployed metropolitan fiber QKD network.

\begin{figure}[hbtp]
    \includegraphics[width=\linewidth, trim=2.15cm 0cm 5.2cm 0cm, clip]{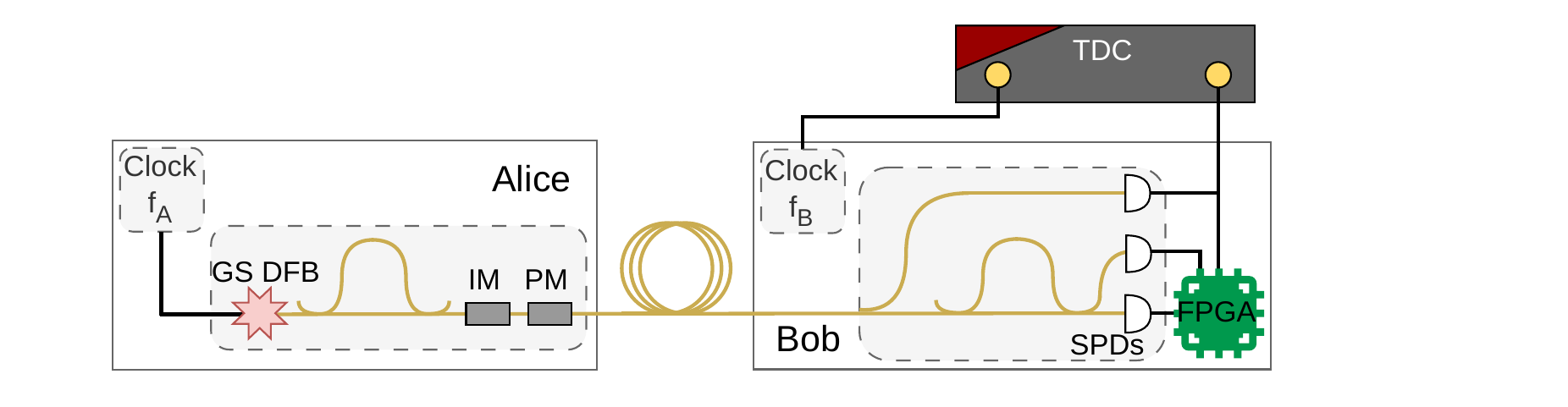}
  \caption{Experimental setup of the QKD systems. GS DFB: gain-switched
distributed feedback laser; IM: intensity modulator; PM: phase modulator; SPD: single-photon detector; FPGA: field-programmable gate array; TDC: time-to-digital converter used for time-correlated single-photon counting (TCSPC) histogram acquisition. Dashed lines represent temperature stabilized regions.}
  \label{fig:setup}
\end{figure}

We focus our study on time-bin BB84\,\cite{bennett1984} QKD systems. In such systems, Alice typically uses an imbalanced interferometer to generate pairs of phase-correlated weak coherent pulses from a pulsed laser, and encodes qubits in either the $Z$ or $X$ basis using intensity and phase modulators\,\cite{boaron2018}. In particular, we consider the commercial system Clavis XG from ID Quantique (Fig.~\ref{fig:setup}). In the $Z$ basis, qubits are encoded by preparing single photons in either an early or a late time bin and are used to encode the raw key. In the $X$ basis, qubits are encoded in the relative phase between the pulses of a pair and are used for error estimation. On the receiver side, after a passive basis choice mechanism, $Z$-basis qubits are detected directly by a single-photon detector (SPD). Meanwhile, $X$-basis qubits interfere within a matching interferometer where two SPDs are used for measurements. 

In standard operation, detections are timestamped by Bob using an internal field-programmable gate array (FPGA), and a service channel distributes a reference signal between Alice's and Bob's clock to prevent a frequency mismatch. Here, we deliberately remove the service channel and access timing information directly from the detections. To this end, Bob records the detection time of each detected photon in the $Z$ basis using an external time-to-digital converter (TDC) to build time-resolved histograms of detection events. The TDC (ID Quantique ID1000) is operated with a start signal coming from Bob’s local clock, running at frequency $f_B$, and a stop signal generated by the SPD output of the $Z$ basis.

Considering that most detections originate from optical pulses emitted by Alice, the stop signal implicitly carries the timing information of Alice’s source, which is driven by her clock at frequency $f_A$. In the absence of active synchronization, Alice’s and Bob’s clocks exhibit a frequency mismatch $\Delta f=f_B-f_A$, which we assume to be constant over the acquisition time of a histogram. This mismatch produces a time shift between Alice’s time measured in Bob’s frame and Bob's time\,\cite{Bregni1997,Hamilton2008}. Indeed, defining the clock drift as
$t_{\text{drift}} \equiv \Delta f/f_A$, the time shift at time \(t\), in Bob's time frame, is
\begin{equation}
    \Delta t(t) = t_{\text{drift}}\,t + t_0\,,
    \label{TER}
\end{equation}
where \(t_0\) is a static offset.
Because the TDC timestamps detections in Bob’s frame, this shift directly affects the measured photon arrival times. Considering Bob's time to be continuous, the expected value of a photon arrival time relative to Bob’s time becomes
\begin{equation}
    \mu_j(t) = t_{\text{drift}}\, t + t_0 + \mu_j\,,
    \label{mu_j(t)}
\end{equation}
where \(\mu_j\) is the nominal center of the early (\(j=e\)) or late (\(j=l\)) time bin. Using the continuous time $t$, and defining $t'$ as the delay between the photon arrival time and Bob's time, the probability density function (PDF) of the photon delay underlying a start--stop histogram acquired over an integration time ${T_\text{int}}$ can be expressed as
\begin{align}
p_j(t')
&= \frac{1}{t_{\text{drift}} T_{\text{int}}}
\Bigg[
\Phi\!\left(\frac{\mu_j + t_0 + t_{\text{drift}} T_{\text{int}} - t'}{\sigma}\right)
\nonumber \\
&\quad
- \Phi\!\left(\frac{\mu_j + t_0 - t'}{\sigma}\right)
\Bigg]\,,
\label{pdf_drift_Tint}
\end{align}
where $\Phi$ is the cumulative distribution function, and $\sigma$ is the standard deviation of the arrival time PDF of a single photon. This expression reveals the dependence of the PDF on the clock drift--induced time shift $\Delta t_\text{drift}(T_\text{int})\equiv t_\text{drift}T_\text{int}$. Consequently, for a given $\sigma$, distinct values of $t_\text{drift}$ and $T_\text{int}$ lead to identical PDFs whenever their product is the same. Note that numerical evaluation is efficient as $t'$ need only span a few standard deviations of $p_j(t')$. 
Notably, The expectation value of $p_j(t')$ is given by 
\begin{equation}
     \mu_{j}^{\left[0,\, T_{\text{int}}\right]}
    = \frac{t_\text{drift}T_\text{int}}{2} + t_0 + \mu_{j}\,.  
\label{eq:u_tot,j}
\end{equation}
$\mu_{j}^{\left[0,\, T_{\text{int}}\right]}$ corresponds to the expected delay of all the early or late photons of a start--stop histogram acquired during $[0,T_\text{int}]$. Moreover, the clock drift can be retrieved independently of any constant time offset using the difference between the expectation values of two different histograms. For example, using two consecutive histograms each acquired over $T_\text{int}$ leads to
\begin{equation}
t_{\text{drift}} 
= \frac{
\mu_{j}^{\left[T_{\text{int}},\, 2T_{\text{int}}\right)} 
- \mu_{j}^{\left[0,\, T_{\text{int}}\right)}}{T_{\text{int}}}\,,
\label{drift_from_means}
\end{equation}
which is the expression that we use in our algorithm. Before introducing the algorithm in detail, we refine our physical model of the photon arrival-time PDF. In particular, we incorporate the detector jitter and analyze how the clock drift--induced time shift changes the QBER. This modeling step is required to determine which clock drift values are acceptable for chosen QKD parameters.

Indeed, Bob's detector adds timing jitter that broadens the photon arrival-time PDF. In the following, we consider a single-photon avalanche diode (SPAD) whose timing response is modeled by a skew normal distribution \(k(\xi, \alpha,\omega)\)\,\cite{Amri2016}. The location parameter $\xi$ is chosen such that the distribution has zero mean. The PDF of the photon detection time relative to Bob's time during an integration time $T_\text{int}$, including SPAD jitter, is given by convolution
\begin{equation}
    p_{j,\text{SPAD}}(t')=(p_j*k)(t')\,.
    \label{eq:folded_pdf_mod}
\end{equation}
Note that the expectation value of $p_{j,\text{SPAD}}(t')$ remains unchanged as $k$ has zero mean. Finally, the total PDF of the photon detection time, assuming an equal proportion of early and late photons, is
\begin{equation}
    p_{\text{tot}}(t')=\frac{1}{2}p_{e,\text{SPAD}}(t')+\frac{1}{2}p_{l,\text{SPAD}}(t')\,.
    \label{eq:ptotspad}
\end{equation}

In practice, histograms can be folded over the duration of one or two time bins for ease of visualization and post-processing. Note that the relative arrival time $t'$ introduced above is an unbounded variable, so that folding needs to be introduced explicitly through a modulo operation. Defining $\tau =t' \bmod{T_\text{hist}}$, where $T_\text{hist}$ is the chosen histogram period, we denote the total folded PDF as $\tilde p_\text{tot}(\tau)$.

The time shift between Alice's and Bob's clocks (Eq.\,\eqref{TER}) induces errors by causing early (late) photons of $\tilde p_\text{tot}(\tau)$ to leak into the late (early) time bin. To isolate the contribution of clock drift to the QBER in the $Z$ basis from that of the static time offset $t_0$, we set $t_0=0$ in Eq.\,\eqref{TER}, set $T_\text{hist}=2T_\text{bin}$, where $T_\text{bin}$ denotes the duration of a time bin, and derive the clock drift--induced error $\epsilon_\text{drift}$ analytically. 
 
Our derivation of $\epsilon_\text{drift}$ accounts for temporal filtering. Indeed, QKD systems can apply temporal filtering windows to the start–stop histograms to reject events that occur far from the expected arrival times, thereby enhancing the signal-to-noise ratio\,\cite{Grunenfelder2023}. For time bin $j$, we define a static filtering time window $T_{j}^{(\text{w})} = [\,\mu_j - w/2,\, \mu_j + w/2)$, centered at the nominal bin center $\mu_j$ and of width $w \in (0,\, T_\text{bin}]$. The probability of an erroneous detection is obtained by integrating the PDF of an early or late photon detection over the static filtering time window associated with the other time bin
\begin{equation}
    P^{(\text{w})}_{j|j'}
= \int_{T_{j'}^{\text{(w)}}} \tilde p_{j,\text{SPAD}}(\tau)\,d\tau\,, \,\,\text{with}\,\,j'\neq j\,.
\label{eq:leaked_prob_mass}
\end{equation}
Following the standard definition of the QBER, the clock drift--induced error evaluated within the filtering windows $\epsilon_\text{drift}^{(\text{w})}$ is defined as the probability of detecting a photon in the incorrect filtering window, normalized by the probability of detecting a photon in either window. Assuming an equal proportion of early and late photons and using 
$P^{(\text{w})}_{j|j'} = P^{(\text{w})}_{j'|j}$, we obtain
\begin{equation}
\epsilon_\text{drift}^{(\text{w})}
= \frac{P^{(\text{w})}_{l|e} + P^{(\text{w})}_{e|l}}
{P^{(\text{w})}_{e|e} + P^{(\text{w})}_{l|e} + P^{(\text{w})}_{e|l} + P^{(\text{w})}_{l|l}} = \frac{P^{(\text{w})}_{j|j'}}
{P^{(\text{w})}_{j'|j'} + P^{(\text{w})}_{j|j'} }\,,
\label{eq:QBER_f}
\end{equation}
which can be solved numerically as a function of $\Delta t_\text{drift}(T_\text{int})$.

The clock drift--induced error $\epsilon_\text{drift}^{(\text{w})}$ represents the contribution of clock drift to the QBER during the time $[0,T_\text{int})$. Starting a histogram acquisition at $t=0$ implies that Alice's and Bob's clocks are initially synchronized, since $\Delta t_\text{drift}(0)=0$. However, starting at a time $t \neq 0$ in Bob’s frame only introduces an additional time offset, which can be taken into account by a simple redefinition of $t_0$, such that all previous equations remain valid. Therefore, the analytical description derived here does not rely on synchronized initial conditions as long as $t_0$ is defined suitably.

\begin{figure}[htbp]
  \centering
  \begin{subfigure}[b]{\linewidth}
    \centering
    \includegraphics[width=\linewidth]{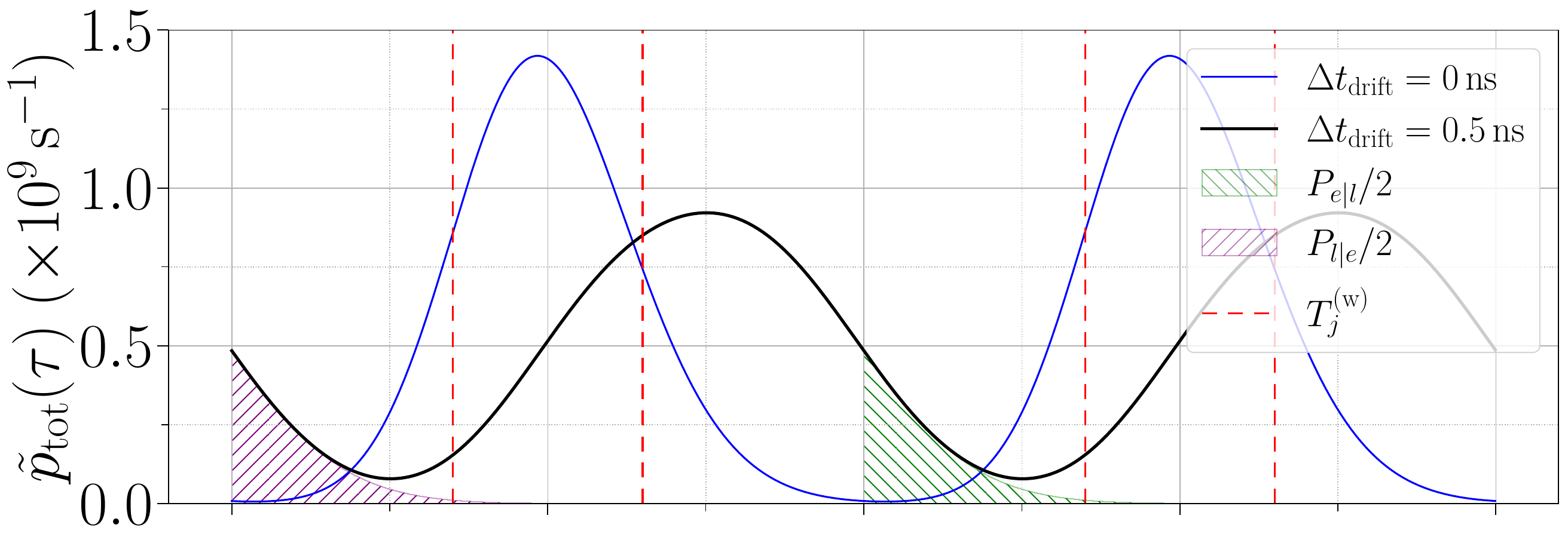}
  \end{subfigure}
  \begin{subfigure}[b]{\linewidth}
    \centering
    \includegraphics[width=\linewidth]{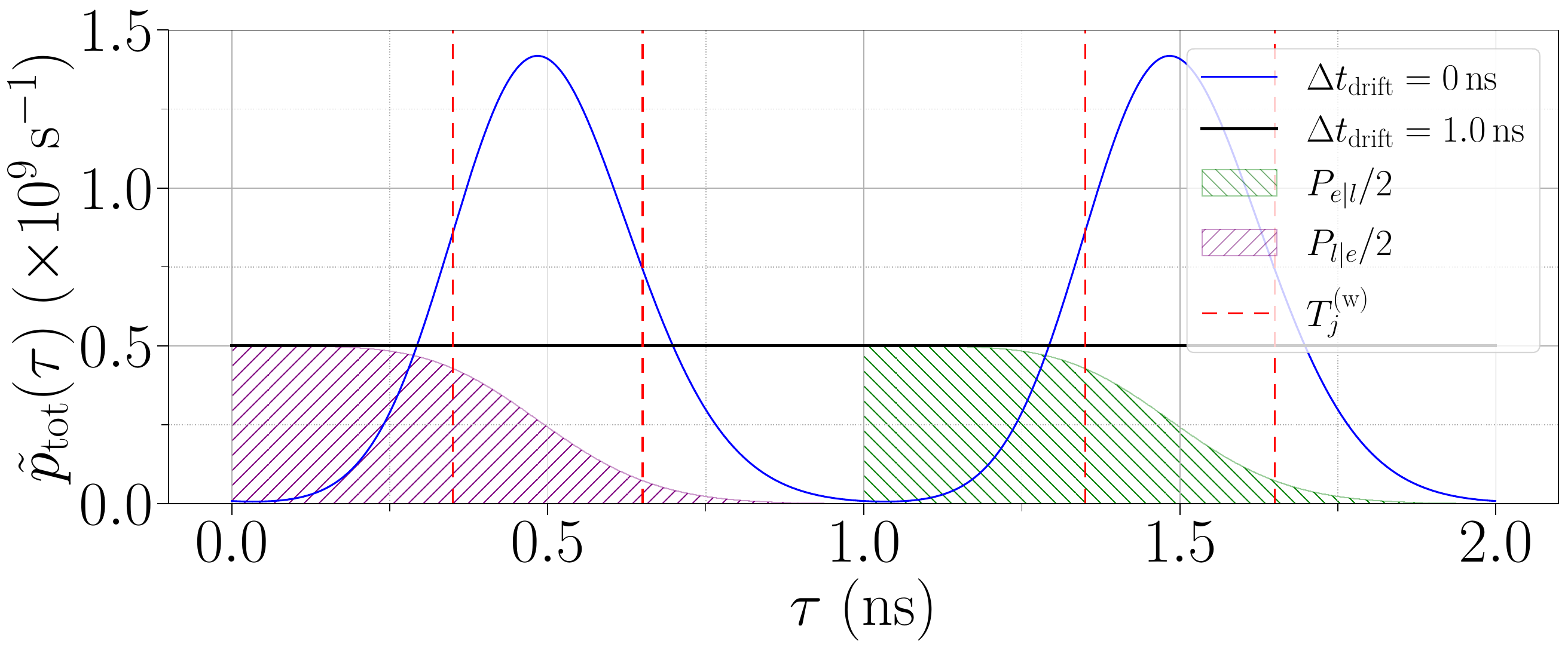}
  \end{subfigure}

  \caption{Probability density functions for \(\Delta t_\text{drift}(T_\text{int})=0.5\,\text{ns}\) (top) and \(1\,\text{ns}\) (bottom). Dashed lines indicate a filtering window of width \(w=300\,\text{ps}\).}
  \label{fig:pdf}
\end{figure}

Fig.~\ref{fig:pdf} shows $\tilde p_{\text{tot}}(\tau)$ for $\Delta t_\text{drift}(T_{\text{int}})=T_\text{bin}/2$ (top) and $T_\text{bin}$ (bottom), with $t_0=0$. To reflect our setup, we consider two time bins of duration $T_\text{bin}=1\,\text{ns}$ centered at \( \mu_e = 0.5\,\text{ns} \) and \( \mu_l = 1.5\,\text{ns} \). The laser wavelength is \(1550\,\text{nm}\) with pulse duration \(\tau_{\text{FWHM}}=77\,\text{ps}\) and quadratic temporal phase \(\beta=-3.7\times10^{20}\,\text{rad}\,\text{s}^{-2}\)\,\cite{Millet2025}. 
The SPAD response is modeled with \(\alpha=3\), \(\omega=150\,\text{ps}\), and the quantum channel length is \(z=120\,\text{km}\). The PDFs are computed from Eq.~\eqref{eq:ptotspad} and errors from Eq.~\eqref{eq:QBER_f}. Under these conditions, $\epsilon_\text{drift}\approx11\,\%$ and $50\,\%$ for $\Delta t_\text{drift}(T_{\text{int}})=0.5\,\text{ns}$ and $1\,\text{ns}$, respectively. A filtering window of width $w=300\,\text{ps}$ reduces the error rate to $\epsilon_\text{drift}^{(\text{w})}=0.4\,\%$ at $0.5\,\text{ns}$, while remaining near $50\,\%$ at $1\,\text{ns}$.

\begin{figure}[htbp]
  \centering
\includegraphics[width=\linewidth]{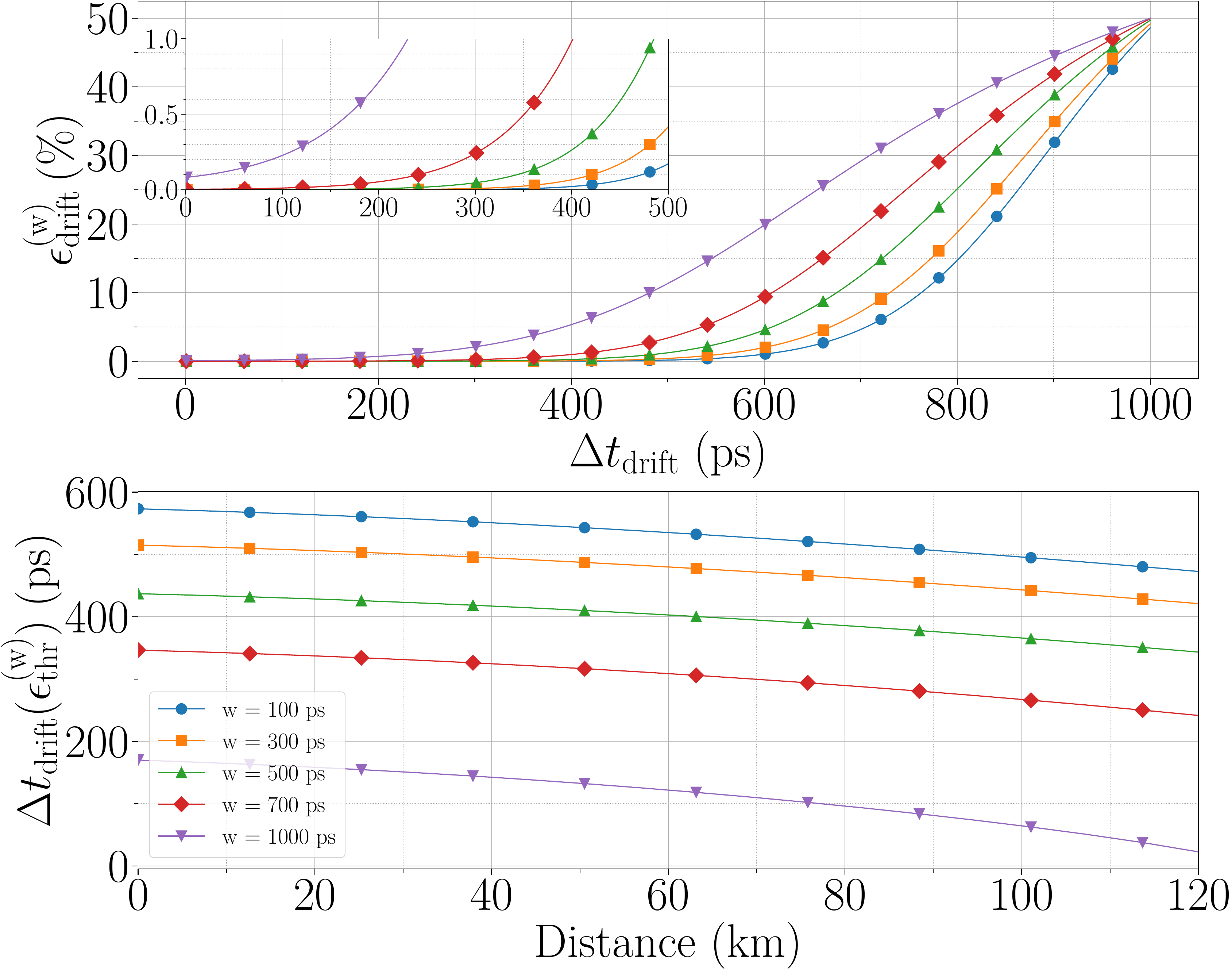}
  \caption{(Top) $\epsilon_\text{drift}^{(\text{w})}$ as a function of $\Delta t_\text{drift}(T_\text{int})$ for different filtering window widths $w$ at $z=120$\,km. (Bottom) $\Delta t_\text{drift}(\epsilon_{\text{thr}}^\text{(w)}=0.1\,\%)$ as a function of the transmission distance $z$ for different filtering window widths $w$.}
  \label{fig:QBER_vs_driftxTint}
\end{figure}

Then, we evaluate Eq.~\eqref{eq:QBER_f} as a function of \(\Delta t_\text{drift}(T_\text{int})\) for different filtering widths and transmission distances, restricting to positive clock drifts (representing worst case due to detector skewness). 
As shown in Fig.~\ref{fig:QBER_vs_driftxTint}, \(\epsilon_\text{drift}^{(\text{w})}\to 0\) for small drift and approaches \(50\,\%\) near \(1\,\text{ns}\). 
Temporal filtering significantly relaxes drift constraints at the expense of reduced detection rate: for \(\epsilon_\text{drift}^{(\text{w})}=0.1\,\%\), the tolerable drift increases from \(\sim 23\,\text{ps}\) (\(w=1000\,\text{ps}\)) to \(\sim 421\,\text{ps}\) (\(w=300\,\text{ps}\)). For \(|\Delta t_\text{drift}(T_{\text{int}})| \gtrsim 1\,\text{ns}\) (not shown), error rate can exceed \(50\,\%\), precluding key generation.

To complete the model, we define \(\Delta t_{\text{drift}}(\epsilon_{\text{thr}}^{(\text{w})})\) as the maximum clock drift such that 
\(\epsilon_\text{drift}^{(\text{w})} \leq \epsilon_{\text{thr}}^{(\text{w})}\). 
As shown in Fig.~\ref{fig:QBER_vs_driftxTint} (bottom), this bound decreases with transmission distance due to chromatic dispersion. It also decreases with increasing window width, as broader windows admit more leaked probability mass.

Here, we propose an algorithm measuring and compensating both the clock frequency mismatch between the clocks of Alice and Bob and fluctuations of the absolute time offset $t_0$. The algorithm is based on Eq.\,\eqref{drift_from_means} and estimates these two quantities from the circular mean of two consecutively measured histograms. Circular means are used to prevent histogram folding artifacts. Bob’s clock frequency and the TDC's delay are then adjusted accordingly.

\vspace{0.5em}
\noindent\textbf{Assumptions.}
We assume that (i) \(t_{\text{drift}}\) and \(t_0\) are constant over \([0,2T_{\text{int}})\), 
(ii) \(|t_{\text{drift}}|T_{\text{int}}<T_{\text{bin}}/2\), and 
(iii) histograms are folded over one time-bin, yielding 
\(\tilde p_\text{tot}(\tau)=\tilde p_{e,\text{SPAD}}(\tau)\) with \(\tau=t'\bmod T_\text{bin}\).

\vspace{0.5em}
\noindent\textbf{Algorithm.}
\begin{enumerate}
\item Acquire a histogram over \([0,T_{\text{int}})\) and estimate the circular mean of $\tau$
\begin{equation}
    \hat m_1=\frac{1}{C}\sum_k c_k\,e^{\frac{2\pi i \tau_k}{T_\text{bin}}}\,,
\end{equation}
where \(\tau_k\) and \(c_k\) denote bin centers and counts, and \(C\) is the total count.
\item Acquire a second histogram over \([T_{\text{int}},2T_{\text{int}})\) and compute \(\hat m_2\).
\item Estimate the signed clock drift as
\begin{equation}
\hat t_{\text{drift}} =\frac{T_{\text{bin}}}{2\pi T_{\text{int}}}
\arg\!\left(\hat m_2\overline{\hat m_1}\right)\,,
\label{tdrift_algo}
\end{equation}
where $\overline{\hat m_1}$ is the complex conjugate of $\hat m_1$. 
\item Estimate the signed delay compensating for the time shift accumulated due to clock drift over $[0,2T_{\text{int}})$
\begin{equation}
\hat\epsilon = \mu_e-\Bigg[\frac{T_{\text{bin}}}{2\pi}\big(\arg(\hat m_1)-\hat\phi_q\big)
+\frac{3}{2}\hat t_{\text{drift}}T_\text{int}\Bigg]\bmod T_{\text{bin}}\,,
\label{epsilon_algo}
\end{equation}
where \(\hat\phi_q\) is estimated from the SPAD distribution.
\item Update the clock frequency as \(f'_B=f_B/(1+\hat t_\text{drift})\).
\item Compensate the time shift by adding $\hat\epsilon$ to the delay of the TDC. As $\hat\epsilon$ is proportional to $t_0 \pmod{T_\text{bin}}$, small fluctuations of the time offset are also compensated. The integer-bin ambiguity of $t_0$ remains unresolved and is outside the scope of the synchronization algorithm.
\item Repeat to track residual variations.
\end{enumerate}

To fully characterize the requirements of a synchronization algorithm, one should also determine the clock-stability conditions under which it can operate reliably. These conditions follow from both the QKD constraints identified by the model above and the limitations of the algorithm.

First, a clock drift such that \(|\Delta t_\text{drift}(T_\text{int})|=\left|t_{\text{drift}}\right|T_{\text{int}}>T_\text{bin}/2\) is indistinguishable from the smaller drift of opposite sign, setting the maximum measurable drift without sign ambiguity. 
This maximum is achieved for the minimum integration time, corresponding to one detected photon per histogram, i.e., \(T_{\text{int}}=1/\text{cps}_{\text{eff,Alice}}\), where \(\text{cps}_{\text{eff,Alice}}\) denotes the detection count rate of Alice's photons corrected for dark counts, dead time, detector efficiency, and losses\,\cite{USMAN2018,Mengler2025}. During the alignment phase of the QKD system (when no keys are generated), the photon number \(\bar{n}\) can be increased, so that the detection rate saturates at \(1/\tau_D\). The absolute maximum measurable drift without sign ambiguity is then
\begin{equation}
    \left|t_{\text{drift,max}}\right|
= \frac{T_\text{bin}}{2\tau_D} \,.
\label{inequality_max_drift}
\end{equation}
If the initial clock drift exceeds this bound, clocks can be calibrated up to a negligible residual drift before usage. After calibration, the residual drift is governed by clock aging. Denoting by \(t_c\) the elapsed time since calibration, the maximum calibration interval \(t_{c,\text{max}}\) is set by
\begin{equation}
    \left|\frac{d\,t_{\text{drift}}}{dt}t_{c,\text{max}}\right| < \left|t_{\text{drift,max}}\right| \,.
    \label{eq:clock_constraint_1}
\end{equation}

Second, maintaining the clock drift--induced error below a threshold $\epsilon^{(\text{w})}_{\text{thr}}$ after the first frequency compensation of the synchronization algorithm requires that the residual clock drift--induced time shift over the next two histogram acquisitions remains smaller than $\Delta t_\text{drift}(\epsilon^{(\text{w})}_{\text{thr}})$. This condition bounds clock stability as
\begin{equation}
\left|\frac{d t_{\text{drift}}}{dt}\right|
\;<\;
\frac{\Delta t_\text{drift}(\epsilon^{\text{(w)}}_{\text{thr}})}{4T_{\text{int}}^2}\,.
  \label{ineq_short_term}
\end{equation}

The previous analysis yields two constraints on the clocks. 
For \(\tau_D=15\,\mu\text{s}\), \(\epsilon^{\text{(w)}}_{\text{thr}}=0.1\,\%\), 
\(w=T_\text{bin}=1\,\text{ns}\), and \(T_\text{int}=500\,\text{ms}\), we obtain 
\(\left|t_{\text{drift,max}}\right|\approx 33\,\mu\text{s}/\text{s}\) and 
\(\Delta t_\text{drift}(\epsilon^{\text{(w)}}_{\text{thr}})/(4T_{\text{int}}^2)\approx23\,\text{ps}/\text{s}^2\). 
As \(\left|t_{\text{drift,max}}\right|\) can be difficult to compensate due to constraints on integration time and photon statistics, we define a practical limit based on realistic operating conditions. Assuming \(\bar{n}=10\) and \(\sim 10\) detected photons per histogram during the alignment phase, we obtain \(T_{\text{int}}\approx155\,\mu\text{s}\), yielding a maximum recoverable drift of \(\sim 3.2\,\mu\text{s}/\text{s}\) from Eq.~\eqref{inequality_max_drift}. Applying a \(70\%\) safety margin to account for Poissonian noise gives a practical limit 
\(\left|t_{\text{drift,max}}^{\text{(practical)}}\right|\approx 2.3\,\mu\text{s}/\text{s}\).

To assess compatibility with low-cost hardware, we compare typical aging values of standard crystal oscillators (XO) with the clock constraints, assuming two identical clocks drifting in opposite directions. For a clock with 1-day aging of \(\sim 500\,\text{ppb}\), the relative drift rate between two of such clocks is \(\left|d t_{\text{drift}}/dt\right|\approx 12\,\text{ps}/\text{s}^2\), consistent with the short-term stability constraint of Eq.~\eqref{ineq_short_term}. 
Over longer timescales, a 10-year aging of \(\sim 50\,\text{ppm}\) yields calibration intervals ranging from \(\sim 0.23\) to \(\sim 3.3\) years, depending on whether practical or theoretical drift limits are considered. Note that the total free-running accuracy of higher-end oscillators can remain within \(\sim 1\,\mu\text{s}/\text{s}\) (1\,ppm) over decades, effectively eliminating the need for calibration.

Next, we demonstrate the synchronization algorithm in (i) a 1-hour laboratory experiment over \(100\,\text{km}\) fiber and with variable attenuation, and (ii) a 24-hour field deployment on a metropolitan network. In both cases, Alice transmits a \(1\,\mu\text{s}\) pseudo-random sequence, and the \(Z\)-basis QBER is evaluated with and without a \(300\,\text{ps}\) filtering window.

The setup (Fig.~\ref{fig:setup}) uses Stratum~3E clocks with free-running accuracy over 20 years of 1\,ppm, \(f_A=500\,\text{MHz}\), SPADs with \(\eta_d=0.25\), \(\tau_D=15\,\mu\text{s}\), and \(1.8\,\text{kHz}\) dark count rate. 
The TDC has \(100\,\text{ps}\) bin width and \(11\,\text{ps}\) delay resolution, corresponding to the typical timing resolution of the Clavis XG when using the internal FPGA. Clocks are initialized with a worst-case drift \(\left|t_{\text{drift,max}}^{\text{(practical)}}\right|\approx2.3\,\mu\text{s}/\text{s}\). The algorithm starts with \(\bar{n}=10\), \(T_{\text{int}}=155\,\mu\text{s}\), and increases \(T_{\text{int}}\) up to \(500\,\text{ms}\), while reducing \(\bar{n}\) to its nominal value. Clock offset recovery is performed during the iteration immediately preceding the first $500\,\text{ms}$ acquisition using Pearson correlation, and the TDC delay is adjusted accordingly. Synchronization from \(\left|t_{\text{drift,max}}^{\text{(practical)}}\right|\) is achieved within \(\sim1.3\,\text{s}\) of histogram acquisition.

\begin{figure}[htbp]
  \centering
  \includegraphics[width=\linewidth]{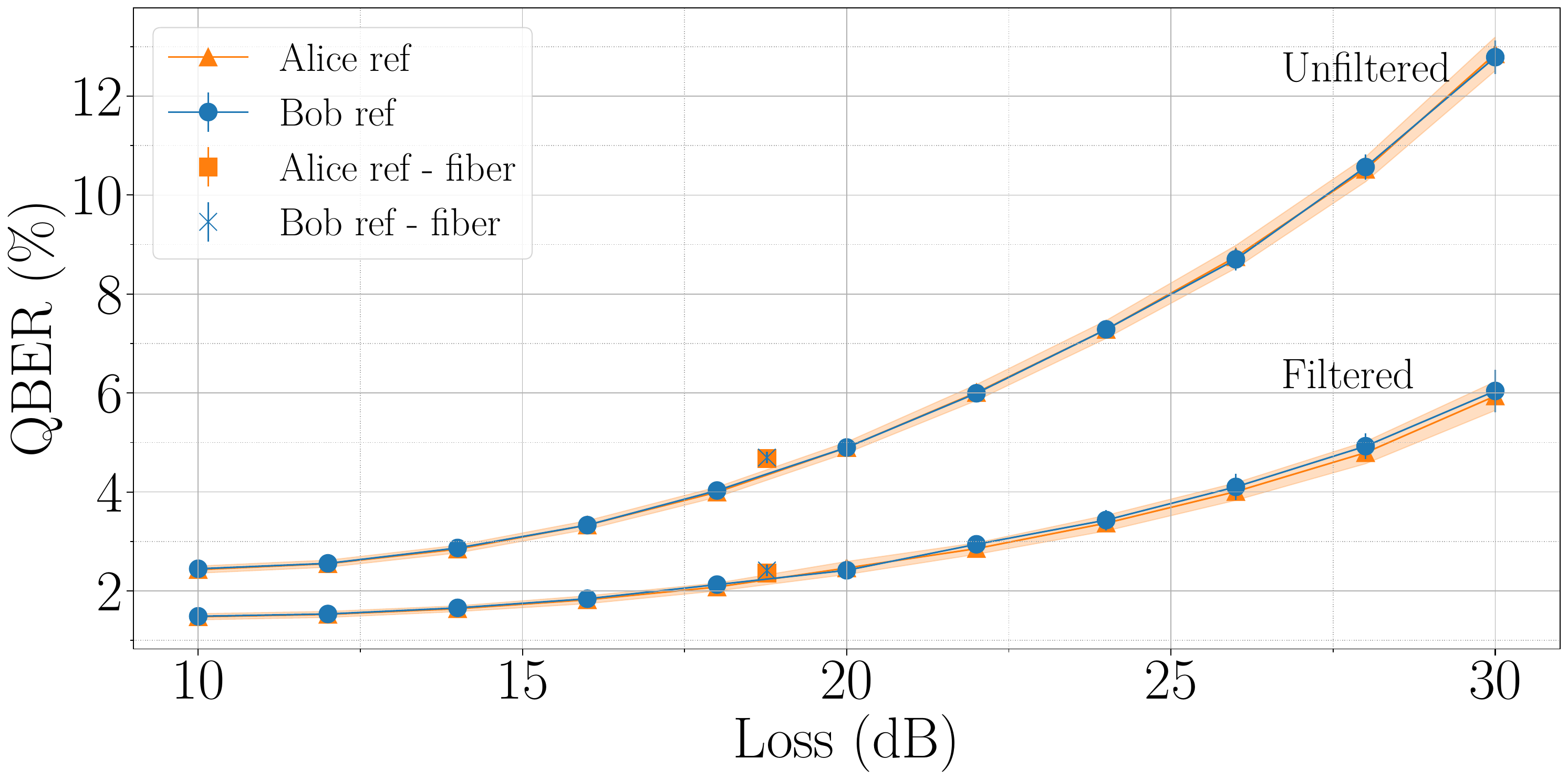}
  \caption{Measured QBER with (\(w=300\,\text{ps}\)) and without temporal filtering as a function of quantum channel loss. Shaded region indicates the standard deviation of the Alice-reference curve.}
  \label{fig:result_voa_fiber}
\end{figure}

Fig.~\ref{fig:result_voa_fiber} shows the QBER versus channel loss (10–30\,dB) after clock offset recovery. Using Bob's clock as the TDC reference, the QBER closely follows the Alice-reference baseline (no clock drift), with deviations \(\sim0.1\,\%\). Fiber measurements have slightly higher QBER due to chromatic dispersion.

\begin{figure}[htbp]
  \centering
  \includegraphics[width=\linewidth]{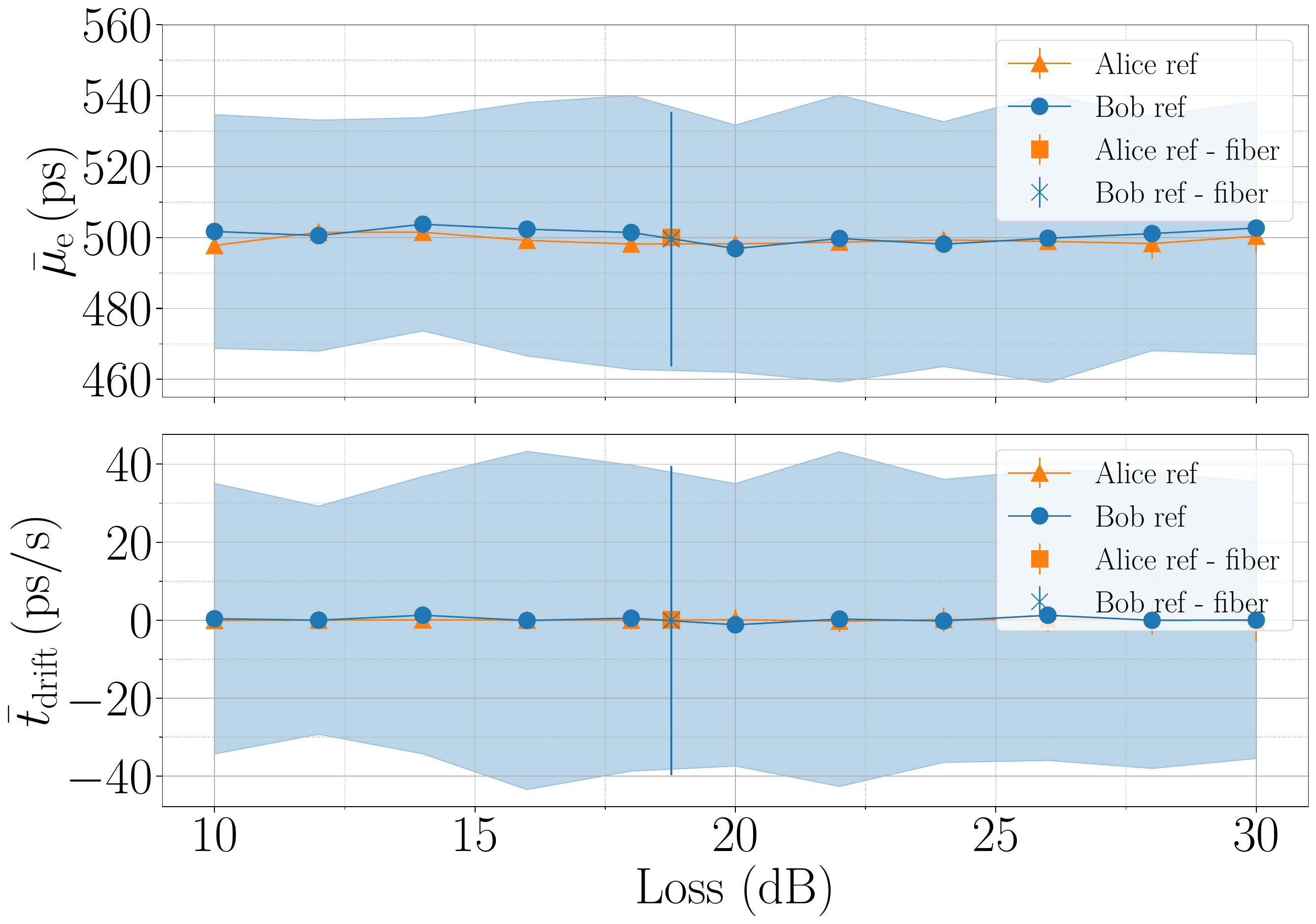}
  \caption{Average detection time (top) and clock drift (bottom) as a function of the quantum channel loss. Shaded region indicates the standard deviation of the Bob-reference curve.}
  \label{fig:result_mean_&drift_voa_fiber}
\end{figure}

Fig.~\ref{fig:result_mean_&drift_voa_fiber} shows the average of photon detection times (Eq.\,\eqref{eq:u_tot,j}) and average clock drift after clock-offset recovery. Results obtained with Bob as the reference follow the baseline, with standard deviations \(\sim40\,\text{ps}\) and \(\sim40\,\text{ps}/\text{s}\), respectively. 

At \(100\,\text{km}\), a clock drift-induced shift of \(80\,\text{ps}\) yields \(\epsilon_{\text{drift}}\sim0.1\,\%\) (see the bottom panel of Fig.~\ref{fig:QBER_vs_driftxTint}), consistent with observations. 
With temporal filtering, the expected error drops to \(\sim10^{-3}\,\%\), while the measured deviation remains \(\sim0.1\,\%\), attributed to discretization of the filtering window.

Finally, we perform field measurements on a section of the Geneva Quantum Network (GQN), interconnecting ID Quantique and two sites of the University of Geneva. The deployed fiber loop spans approximately $16\,\text{km}$ and exhibits a total optical loss of approximately $11.5\,\text{dB}$. 

\begin{figure}[htbp]
  \centering
  \includegraphics[width=\linewidth]{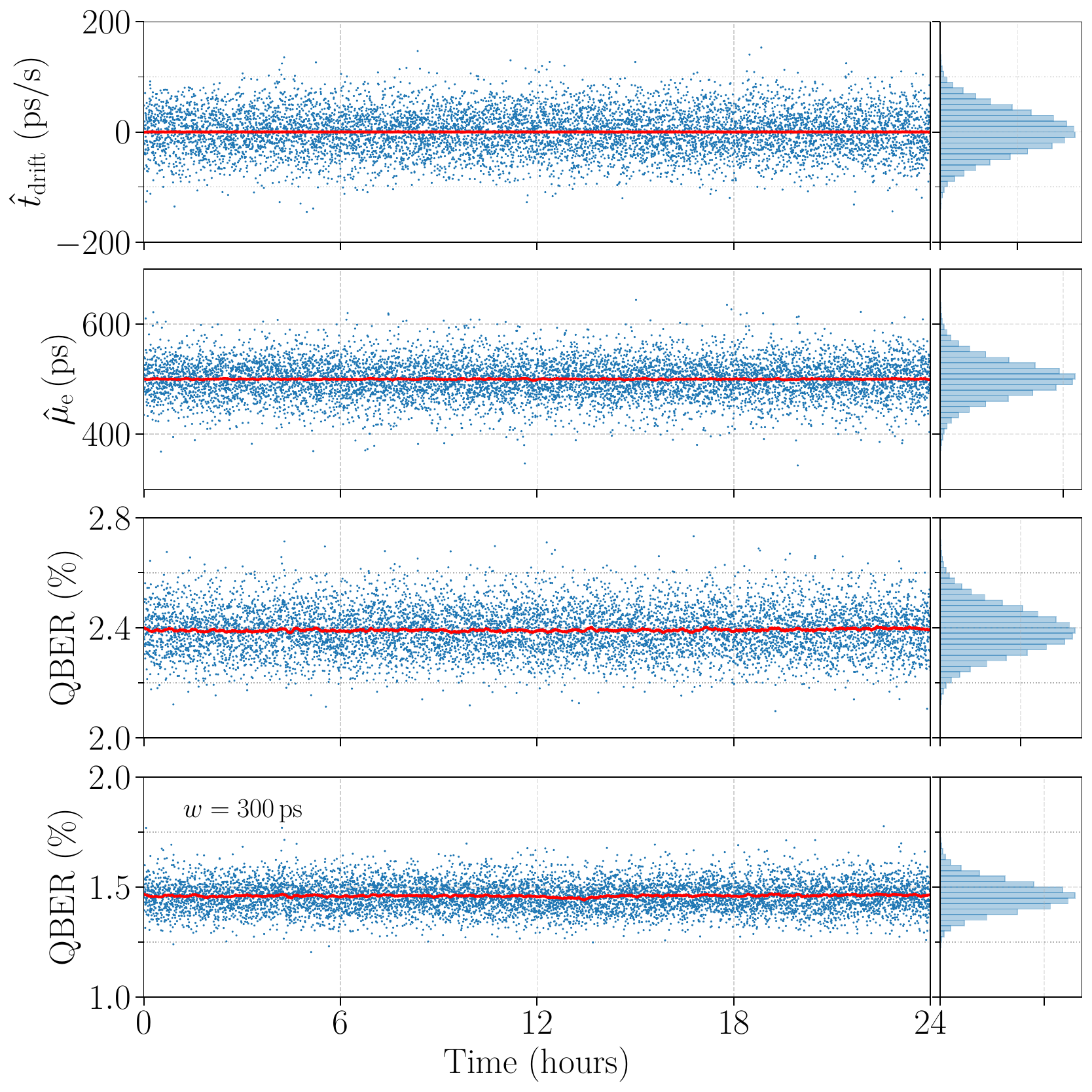}
  \caption{Clock drift $\hat t_\text{drift}$, mean photon detection time $\hat \mu_\text{e}$, QBER, and filtered QBER with $w=300\,\text{ps}$ after clock offset recovery. Solid lines represent 10-minute moving averages. Each histogram is normalized to its maximum value.}
  \label{fig:GQN_result_time}
\end{figure}

Results on the GQN are shown in Fig.~\ref{fig:GQN_result_time} and demonstrate stable operation over 24 hours. After time-offset recovery, the average photon detection time is \(500.0\pm34.3\,\text{ps}\), the residual clock drift is \(0.0\pm39.4\,\text{ps}/\text{s}\), the QBER is \(2.39\pm0.0836\,\%\), and reduces to \(1.46\pm0.0671\,\%\) with temporal filtering. The time deviation\,\cite{Riley2008} (TDEV) remains \(\sim24\,\text{ps}\), indicating a stability floor set by timing resolution and residual clock noise.

Clock synchronization in fiber-based time-bin QKD reduces to a simple effect: clock drift reshapes start--stop histograms through the accumulated time shift $\Delta t_{\text{drift}}(T_{\text{int}})$, with errors scaling accordingly. This enables a lightweight synchronization algorithm based on two consecutive folded histograms, where a circular-mean estimate yields the signed drift for frequency correction, and a delay update compensates for time offsets without any dedicated synchronization channel. An analytical model linking clock drift-induced time shifts to errors provides explicit clock-stability requirements and practical design rules. Experiments in laboratory and deployed networks demonstrate stable operation with performance comparable to an ideal clock reference. Beyond QKD, this approach may extend to other time-tagged quantum communication and sensing systems requiring synchronization directly from detection statistics\,\cite{Komar2014, Yin2017, Wehner2018}.

\begin{acknowledgments}
The authors thank Raphaël Houlmann for helpful discussions. The authors acknowledge financial support from the Marie Sklodowska-Curie Grant No. 101072637 (Project Quantum-Safe-Internet) and the Swiss State Secretariat for Research and Innovation (SERI)(Contract No. UeM019-3).
\end{acknowledgments}

\appendix
\section{Expected arrival time of a photon}
In this section, we derive the expected photon arrival time relative to Bob’s time (Eq.~\eqref{mu_j(t)}).
\subsection{Clock drift definition and clock drift--induced time shift}
\label{subsec:clock_drift}
In many time-synchronized systems, events are time-stamped using local clocks that may run at different frequencies. Even a small mismatch between two clocks causes their time to diverge, producing a linear shift in the measured times of shared events. Let Bob’s clock frequency be
\begin{equation}
    f_B = f_A + \Delta f\, ,
\end{equation}
where \(f_A\) is Alice’s frequency and \(\Delta f\) is a constant deviation. After \(N\) clock cycles, the elapsed times in Alice’s and Bob’s frames are
\begin{equation}
       t_A = \frac{N}{f_A}\,, 
    \qquad
    t_B = \frac{N}{f_B}\,, 
\end{equation}
and Alice's time measured in Bob's time frame is
\begin{equation}
   t^{(m)} = \frac{N}{f_A} + t_0= \frac{N}{f_B - \Delta f} + t_0\,, 
\end{equation}
where $t_0$ is the time offset between Alice's and Bob's time at $N=0$. The deviation between Alice's time measured by Bob and Bob's time, is
\begin{align}
\Delta t 
    &= t^{(m)} - t_B
     = \frac{N}{f_A}+t_0 - \frac{N}{f_B} 
     = N\,\frac{f_B - f_A}{f_A f_B}+t_0\,,
\end{align}
since Bob’s elapsed time is $t_B = {N}/{f_B}$, and defining the clock drift (fractional frequency offset) as 
\begin{equation}
    t_{\text{drift}} 
    = \frac{f_B - f_A}{f_A}\,,
\end{equation}
we obtain the linear drift model
\begin{align}
    \Delta t(t_B) =t_\text{drift}t_B + t_0\,.
    \label{linear_drift_model}
\end{align}

\subsection{Expected arrival time of a photon}
\label{subsec:Mean_arrival_time_of_a_photon}
Our goal is to derive an expression of the expected arrival time of a detected photon as a function of the clock drift. Let us consider that Alice emits the photon $N$ at the $N^\text{th}$ clock period. 
The emission time in Alice's frame is
\begin{equation}
    t_A = \frac{N}{f_A} + \delta_j\Delta_\text{IF}\,,
\end{equation}
where $\Delta_\text{IF}$ is the interferometer delay between the early and late paths, and $\delta_j$ equals $0$ for early photons ($j=e$) and $1$ for late photons ($j=l$). Meanwhile, Bob's local time for the same index $N$ is $t_B = N/f_B$ with $f_B = f_A + \Delta f$. Bob timestamps the detections using his own clock, so the measured arrival time is
\begin{equation}
t^{(m)} = \frac{N}{f_A} + t_0 +  \delta_j\Delta_\text{IF} + \epsilon\,,
\end{equation}
where $t_0$ includes both the initial offset between Alice's and Bob's clocks and the photon time-of-flight, and $\epsilon$ is the detector timing jitter with zero mean. It is convenient to write the measured arrival time such that photons arrive at the center of their time bin 
\begin{align}
t^{(m)}
&= \frac{N}{f_A} + t_0 - \frac{T_\text{bin}}{2} + \frac{T_\text{bin}}{2} + \delta_j\Delta_\text{IF} +\epsilon \nonumber\\
&= \frac{N}{f_A} + t_0' + \mu_j + \epsilon\,,
\end{align}
where $T_\text{bin}$ represents the duration of a time bin, $\Delta_\text{IF}=T_\text{bin}$, $\mu_j\equiv \frac{T_\text{bin}}{2} + \delta_j\Delta_\text{IF}$ is the center of the early ($j=e$) or late ($j=l$) time bin, $t_0'\equiv t_0 - T_\text{bin}/2$, and henceforth $t_0'$ is denoted as $t_0$ for simplicity.

In practice, only a small fraction of photons emitted by Alice are actually detected by Bob. The ideal detection rate due to Alice’s photons is
\begin{equation}
    \text{cps}_{\text{Alice}} = f_A \,\bar{n}\, \eta_d \,\eta_{ch}\,,
\end{equation}
where $\bar{n}$ is the mean photon number per pulse, $\eta_d$ is the photon detection efficiency of Bob's detector, and $\eta_{ch}$ is the transmittance of the quantum channel. The ideal total detection rate is obtained by adding the detector dark count rate (DCR)
\begin{equation}
    \text{cps}_{\text{ideal,tot}}  = \text{cps}_{\text{Alice}} + \text{DCR}\,.
\end{equation}
However, because of the detector dead time $\tau_d$, the effective rate of detections due to Alice's photons is
\begin{align}
\text{cps}_{\text{eff,Alice}}
  &= \frac{\text{cps}_{\text{Alice}}}{
     1 + \text{cps}_{\text{ideal,tot}}\,\tau_d}\,,
\label{eq:cps_eff_Alice}
\end{align}
the effective dark count rate is
\begin{align}
\text{cps}_{\text{eff,dc}}
  &= \frac{\text{DCR}}{
     1 + \text{cps}_{\text{ideal,tot}}\,\tau_d}\,,
\label{eq:cps_eff_DC}
\end{align}
and the average time interval between two successive detections originating from Alice's photons is ${1}/{\text{cps}_{\text{eff,Alice}}}$. Since detections occur irregularly in terms of emission index $N$ but can be expressed regularly in terms of Bob’s detection index $n$, it is natural to label timestamps in terms of Bob’s detection index $n$. 

Indeed, the $n^\text{th}$ detected photon corresponds to some random emission index $N_n$ in Alice’s pulse train. Defining the random variable $N_n$ as the emission index of the $n^\text{th}$ detected photon, the arrival time of the $n^\text{th}$ detected photon measured by Bob $t_n^{(m)}$ is
\begin{equation}
    t_n^{(m)}\equiv t^{(m)}(N_n)=\frac{N_n}{f_A}+t_0+\mu_j+\epsilon\,.
    \label{eq:exact_detection_time}
\end{equation}
Defining $\alpha$ such that $\text{cps}_{\text{eff,Alice}}\equiv\alpha f_A$, $\alpha$ represents the mean fraction of emitted signal photons that result in detections, after correction for dark counts, dead time, detector efficiency, and losses. For the purpose of estimating the expectation value of $N_n$, we introduce an effective model in which each emitted photon leads to a detection with probability $p=\alpha$. Note that this effective description is used only as a mean-field model to obtain the expectation value of $N_n$ and does not imply such statistics for the underlying detection process. In this effective model, let $X_k \in \{0,1\}$ denote a random variable indicating whether the $k^{\text{th}}$ emitted pulse leads to a detection, with $\Pr(X_k = 1) = \alpha$. The total number of detected photons after $N$ emitted pulses is
\begin{equation}
S_N = \sum_{k=1}^{N} X_k\,.
\end{equation}
The emission index $N_n$ of the $n^{\text{th}}$ detected photon is defined as the smallest integer such that $S_{N_n}=n$. Within this effective Bernoulli model, $N_n$ follows a negative-binomial distribution, whose expectation value is
\begin{equation}
\mathbb{E}[N_n] = \frac{n}{\alpha}\,,
\end{equation}
such that the expectation value of $t_n^{(m)}$ is
\begin{equation}
    \mathbb{E}[t_n^{(m)}]=\mathbb{E}\left[\frac{N_n}{f_A}+t_0+\mu_j+\epsilon\right]=\frac{n}{\alpha f_A}+t_0+\mu_j\,.
\end{equation}
The expected deviation between the arrival time of the $n^\text{th}$ detected photon measured by Bob $t_n^{(m)}$ and Bob's time at the detection index $n$ is 
\begin{align}
       \mathbb{E}\left[\Delta t (t_B)\right]
       &= \mathbb{E}[t_n^{(m)} - t_B] \nonumber \\
       &= \frac{n}{\alpha f_A} + t_0 +\mu_j - \frac{n}{\alpha f_B} \nonumber\\
       &= \frac{f_B - f_A}{f_A}\,\frac{n}{\alpha f_B} + t_0 +\mu_j \nonumber\\
       &= t_{\text{drift}}\mathbb{E}[t_B] + t_0 +\mu_j\,.
\label{eq:mean-vs-n}
\end{align}
Note that $\Delta t_{\text{avg}}\equiv\mathbb{E}[t_B]/n$ represents the expected time interval between two successive detections in the absence of a frequency mismatch. The expectation value of the arrival time of the $n^\text{th}$ detected photon measured by Bob is
\begin{equation}
\mathbb{E}[t_n^{(m)}]=\mathbb{E}\left[\Delta t(t_B)+t_B\right]=t_{\text{drift}}n\Delta t_{\text{avg}} + n\Delta t_{\text{avg}} + t_0 +\mu_j \,.
\end{equation}

The discrete index $n$ can be mapped to Bob’s continuous elapsed time through $t = n\,\Delta t_\text{avg}$. Substituting this relation into the expectation value above, and subtracting Bob's time, yields a continuous mean arrival time function 
\begin{equation}
    \mu_j(t) \equiv \mathbb{E}[t_n^{(m)}] - t
    = t_\text{drift}\, t + t_0 + \mu_j\,.
\end{equation}

Subtracting $t$ corresponds to expressing Alice’s photon arrival time relative to Bob’s clock, as performed by a start–stop TDC measurement in which the start is defined by Bob’s clock. In the absence of clock drift ($t_\text{drift} = 0$) and after removing the static offset ($t_0 = 0$), Alice's time would align perfectly with Bob’s reference, so that $\mu_j(t) = \mu_j$, a constant corresponding to the center of the time bin. A nonzero drift produces a linear deviation with slope $t_\text{drift}$, causing the mean arrival time to move away from the nominal bin center $\mu_j$ during the integration period. Thus, $\mu_j(t)$ captures the drift-induced deviation of Alice’s time relative to Bob’s in the mean arrival time of the detected photons, which is what is observed in a start--stop histogram.

\section{Probability density function of photon delay in a start-stop histogram}
In this section, we derive the probability density function (PDF) of the relative arrival time of photons acquired in a start-stop histogram (Eq.~\eqref{pdf_drift_Tint}), as well as its expectation value and standard deviation.
\subsection{Probability density function of the photon arrival time in the presence of clock drift}
\label{subsec:laser}
The electric field of a laser pulse can be represented as $E(t)=\sqrt{I(t)}\cos\left(\omega_{0}t+\phi(t)\right)$. $\omega_0$ is the angular frequency of the pulse carrier, $\phi(t)$ is a phase dependent on time, $\sqrt{I(t)}=\exp(-{t}^2/{\text{t}_g}^2)$, $t_g = {\tau_{\text{FWHM}}}/\sqrt{2 \ln 2}$, and $\tau_{\text{FWHM}}$ is the pulse duration (FWHM of intensity). For a distributed feedback laser, the time-dependent phase can be approximated by a quadratic term\,\cite{Millet2025} $\phi(t)=\beta t^2$.
During propagation in a fiber, the pulse will spread in time as a function of the distance $z$, and the resulting electric field is proportional to
\begin{equation}
E(t,z)\propto \exp\left[\frac{-\tau^2}{t_g(z)^2}(1-i\xi z)\right]\,,
\end{equation}
with  
\begin{equation}
t_g(z)^2=\frac{t_g^2\left(1+\xi^2z^2\right)}{\left(1+\beta^2t_g^4\right)}\,,
\end{equation}
and
\begin{equation}
\xi=\frac{2k^"\left(1+\beta^2t_g^4\right)-\frac{\beta t_g^4}{z}}{t_g^2}\,.
\end{equation}
$k^"$ is the group velocity dispersion and is expressed as $k^"=-\frac{\lambda ^2}{2\pi\text{c}}D$. $D$ is the fiber dispersion coefficient, and $D\approx17\,\text{ps}/\text{km}/\text{nm}$ for standard SMF fiber at $1550\,\text{nm}$. $\tau=t-z/v_g(\omega_0)$ with $v_g$ the phase velocity. Notably, the intensity profile is proportional to
\begin{equation}
I(t,z)\propto \exp\left[\frac{-2\tau^2}{t_g(z)^2}\right]\,.
\end{equation}

For our purpose, the time of arrival of a single photon $t_n^{(m)}$ can be modeled as a Gaussian random variable $\mathcal N(\mathbb{E}[t_n^{(m)}],\sigma^2)$ whose probability density function (PDF) is determined by the laser pulse intensity profile. The intensity profile can be rewritten as
\begin{equation}
    I(t_n^{(m)},\,z) \propto \exp\left[ \frac{-2 \left(t_n^{(m)}-\mathbb{E}[t_n^{(m)}]\right)^2}{t_g(z)^2} \right]\,,
\end{equation}
where $t_g(z)$ denotes the pulse width at the distance $z$ and takes into account the chromatic dispersion in the quantum channel. Identifying the intensity profile with a Gaussian distribution yields the photon time-of-arrival probability density function
\begin{equation}
    \varphi_j(t_n^{(m)})= \frac{1}{\sqrt{2\pi}\sigma}
    \exp\left[-\frac{ \left(t_n^{(m)}-\mathbb{E}[t_n^{(m)}]\right)^2}{2\sigma^2} \right]\,,
\end{equation}
where $\sigma=t_g(z)/2$ is the standard deviation of the distribution, and $j$ accounts for an early or late photon. If one would like to express the photon arrival time relative to Bob's time $t'\equiv t_n^{(m)}-t$, as done in a start--stop histogram, we get

\begin{equation}
    \varphi_j(t'+t)=\frac{1}{\sqrt{2\pi}\sigma}\exp\left[-\frac{\left(t'-\mu_j(t)\right)^2}{2\sigma^2}\right]=\mathcal \varphi_j(t';t)\,.
\end{equation}

\subsection{Probability density function of photon delay in a start-stop histogram}
\label{subsec:exp_value_pdf}

The PDF of the photon relative arrival time underlying a start--stop histogram acquired over an integration time ${T_\text{int}}$ is the sum of the individual PDFs of the $T_\text{int}/\Delta t_\text{avg}$ photons normalized by the number of photons (a Gaussian mixture of equal weights). Using the continuous time $t$, this corresponds to the normalized integral of the individual PDF $\mathcal \varphi_j(t';t)$ over ${T_\text{int}}$
\begin{equation}
    p_j(t')
    = \frac{1}{T_\text{int}}\int_{0}^{T_\text{int}}\varphi_j(t';t)\,dt\,,
\end{equation}
which can  be rewritten as
\begin{align}
    p_j(t')
    &=\frac{1}{t_\text{drift}T_\text{int}}\int\limits_{\mu_j+t_0}^{\mu_j+t_0+t_\text{drift} T_\text{int}}{ \varphi_j(\mu_j(t);t')\,d\mu_j(t)} \nonumber\\
    &= \frac{1}{t_{\text{drift}} T_{\text{int}}}
\Bigg[
\Phi\!\left(\frac{\mu_j + t_0 + t_{\text{drift}} T_{\text{int}} - t'}{\sigma}\right)
\nonumber \\
&\quad
- \Phi\!\left(\frac{\mu_j + t_0 - t'}{\sigma}\right)
\Bigg]\,,
\end{align}
where $\Phi$ is the cumulative distribution function. The expectation value of $p_j(t')$ is given by
\begin{align}
     \mu_{j}^{\left[0,\, T_{\text{int}}\right]} \nonumber
     &=\int_{-\infty}^{\infty} t' \, p_j(t')\,dt'\\
    &= \frac{t_\text{drift}T_\text{int}}{2} + t_0 + \mu_{j}\,,   
\end{align}
and the standard deviation of $p_j(t')$ is given by 
\begin{equation}
\sigma_\text{tot}=\sqrt{\sigma^2+\operatorname{Var}\left[\mu_j(t)\right]}=\sqrt{\sigma^2+\frac{\left(t_\text{drift}T_\text {int}\right)^2}{12}}\,.
\end{equation}
For completeness, note that the standard deviation of $p_{j,\text{SPAD}}(t')$ is given by
\begin{equation}
    \sigma_\text{tot,SPAD} = \sqrt{\sigma_\text{tot}^2+\sigma_\text{SPAD}^2}=\sqrt{\sigma_\text{tot}^2+\omega^2(1-\frac{2\delta^2}{\pi})}\,,
\end{equation}
where $\delta=\alpha/\sqrt{1+\alpha^2}$.

\subsection{Histogram folding}
\label{subsec:histogram_folding}
As explained in the main text, the start-stop delay $t'$ is an unbounded variable, and folding needs to be introduced explicitly through a modulo operation. If Bob’s clock period is an integer multiple of the histogram folding period $T_\text{hist}$,
\begin{equation}
\frac{1}{f_B} = kT_{\text{hist}}\,, \qquad k \in \mathbb{N}\,,
\end{equation}
expressing photon arrival times relative to Bob’s clock using $t'$ does not alter the subsequent folding modulo $T_{\text{hist}}$. Indeed, we have
\begin{align}
    t'\pmod {T_{\text{hist}}}
    &= t_n^{(m)}-t_B\pmod {T_{\text{hist}}} \nonumber\\
    &= \Big(t_n^{(m)}\bmod T_{\text{hist}}
    \nonumber \\
    &\quad - \big(\tfrac{N_n}{f_B}\bmod T_{\text{hist}}\big)\Big)
    \pmod {T_{\text{hist}}} \nonumber\\
    &= t_n^{(m)} \pmod {T_{\text{hist}}}\,.
\end{align}

\section{Numerical integration of the probability density function}
\label{sec:num_integration}

For numerical integration, the folded PDF $\tilde p_{j,\text{SPAD}}(\tau)$ can be constructed from the SPAD-convolved distribution $p_{j,\text{SPAD}}(t')$ by summing over all integer multiples of the histogram period $T_\text{hist}= 2T_{\text{bin}}$, with $\tau= t' \bmod{T_\text{hist}}$ and $t'$ is the unbounded time
\begin{equation}
\tilde p_{j,\text{SPAD}}(\tau)
    = \sum_{m\in\mathbb{Z}}
      p_{j,\text{SPAD}}\!\bigl(\tau+m\,T_{\text{hist}}\bigr)\,.
\label{eq:folded_pdf}
\end{equation}
In practice, one can use a finite integer sum, such that the leakage probability into the incorrect time window $T_{j'}^\text{(w)}$ is
\begin{equation}
P_{j|j'}^\text{(f)}= \int_{T_{j'}^\text{(w)}}\sum_{m=m_{\text{min}}}^{m_{\text{max}}} 
p_{j,\text{SPAD}}\!\bigl(\tau + m\,T_{\text{hist}}\bigr)\,d\tau\,,
\label{eq:Pl_e_msum_clean}
\end{equation}
and the integer bounds $(m_{\text{min}},m_{\text{max}})$ are chosen such that the PDF tails are negligible outside that range.  
The function $p_{j,\text{SPAD}}(t')$ can conveniently be written as 
\begin{align}
p_{j,\text{SPAD}}(t')
&= \int_{-\infty}^{\infty} p_j(t'-u)\,k(u)\,du \nonumber\\
&\hspace{-2em}= \frac{1}{t_{\text{drift}}T_{\text{int}}}\int_{-\infty}^{\infty}\!
\Biggl[
\Phi\!\left(\frac{\mu_j+t_0+t_{\text{drift}}T_{\text{int}}-t'+u}{\sigma}\right)
\nonumber\\
&\hspace{-2em}
-\Phi\!\left(\frac{\mu_j+t_0-t'+u}{\sigma}\right)
\Biggr] k(u)\,du \,,
\label{eq:p_spad_unwrapped}
\end{align}
where $\Phi$ is the cumulative distribution function of the standard normal distribution. Substituting Eq.\,\eqref{eq:p_spad_unwrapped} into Eq.\,\eqref{eq:Pl_e_msum_clean} and setting $t_0=0$ yields an expression that can be used for numerical integration.

\section{Clock synchronization algorithm}
\label{subsec:clock_sync_algo}
In this section, we derive Eq.\,\eqref{tdrift_algo} and Eq.\,\eqref{epsilon_algo} of the synchronization algorithm. 

\subsection{PDF for a folded histogram during the synchronization algorithm}
\label{subsec:pdf_folded_hist_during_sync}
The PDF of the photon detection time during an acquisition of duration $T_\text{int}$ can be rewritten from Eq.\,\eqref{eq:ptotspad} as 
\begin{equation}
p_\text{tot}(t')=\frac{1}{2}p_\text{e,SPAD}(t')+\frac{1}{2}p_\text{e,SPAD}(t'-T_\text{bin})\,.
\label{eq:two_lobes_model}
\end{equation}
Defining the histogram folding period $T_\text{hist}=T_\text{bin}$ and $\tau = t' \bmod{T_\text{bin}}$, where $t'$ denotes the unbounded start--stop delay, the folded PDF during the synchronization algorithm is
\begin{align}
\tilde p_{\text{tot}}(\tau)
&= \frac{1}{2}\sum_{m\in\mathbb{Z}} p_{\text{e,SPAD}}(\tau+mT_\text{bin})
\nonumber \\
&\quad + \frac{1}{2}\sum_{m\in\mathbb{Z}} p_{\text{e,SPAD}}(\tau+(m-1)T_\text{bin}) \nonumber\\
&= \tilde p_{\text{e,SPAD}}(\tau)\,.
\label{eq:pdf_e_folded}
\end{align}

\subsection{Recovering the clock drift from a circular mean}
Over an acquisition duration \([0,T_{\text{int}})\), the detection time PDF can be written as 
\begin{equation}
p_{e,\text{SPAD}}(t')=q\!\bigl(t' - \mu_{e}^{[0,T_\text{int})}\bigr)\,,
\label{eq:loc_family}
\end{equation}
where the function \(q\) can be identified from Eq.\,\eqref{eq:p_spad_unwrapped} as
\begin{equation}
\begin{aligned}
q(t')
&= \frac{1}{t_{\text{drift}}T_{\text{int}}}\int_{-\infty}^{\infty}\!
\Biggl[
\Phi\!\left(\frac{t_{\text{drift}}T_{\text{int}}/2 - t' + u}{\sigma}\right)
 \\
&\quad
-\Phi\!\left(\frac{-t_{\text{drift}}T_{\text{int}}/2 - t' + u}{\sigma}\right)
\Biggr] k(u)\,du\,,
\end{aligned}
\label{eq:q(t')}
\end{equation}

and where $\mu_{e}^{[0,T_\text{int})}$ is the expected photon arrival time relative to Bob's (Eq.\,\eqref{eq:u_tot,j}). Histogram folding prevents the use of simple weighted averages to estimate Eq.\,\eqref{drift_from_means}, while circular means remain well defined. The circular mean of the first histogram is
\begin{align}
    m_1
    &=\int e^{\frac{2\pi i t'}{T_\text{bin}}}q(t'-\mu_{e}^{[0,T_\text{int})})\,dt'\nonumber\\
    &=\int e^{\frac{2\pi i (u+\mu_{e}^{[0,T_\text{int})})}{T_\text{bin}}}q(u)\,du \nonumber\\
    &=e^{\frac{2\pi i \mu_{e}^{[0,T_\text{int})}}{T_\text{bin}}}\int e^{\frac{2\pi i u}{T_\text{bin}}}q(u)\,du\,.
\end{align}

The circular mean of the second histogram is
\begin{align}
    m_2
    &=\int e^{\frac{2\pi i t'}{T_\text{bin}}}q(t'-\mu_{e}^{[T_\text{int},2T_\text{int})})\,dt'\nonumber \\
    &=e^{\frac{2\pi i \mu_{e}^{[T_\text{int},2T_\text{int})}}{T_\text{bin}}}\int e^{\frac{2\pi i u}{T_\text{bin}}}q(u)\,du\,.
\end{align}
To isolate the clock drift--induced time shift, we can multiply by the complex conjugate $\overline{m_1}$
\begin{equation}
   m_2\overline{m_1}
= e^{\frac{2\pi i}{T_\text{bin}}\bigl(\mu_{e}^{[T_\text{int},2T_\text{int})}-\mu_{e}^{[0,T_\text{int})}\bigr)}\,|C|^2\,,
\end{equation}
with $C\equiv\int e^{\frac{2\pi i u}{T_\text{bin}}}q(u)\,du$.
Therefore,
\begin{equation}
\begin{aligned}
\arg\!\left(m_2\overline{m_1}\right)
&= \frac{2\pi}{T_{\text{bin}}}
\Bigl(\mu_{e}^{[T_\text{int},2T_\text{int})}
- \mu_{e}^{[0,T_\text{int})}\Bigr) \\
&= \frac{2\pi}{T_{\text{bin}}}\,\Delta t_{\text{drift}}
\quad (\bmod 2\pi)\,.
\end{aligned}
\end{equation}
Assuming the clock drift--induced time shift obeys
\begin{equation}
\left|\Delta t_{\text{drift}}\right| < \frac{T_{\text{bin}}}{2},
\end{equation}
and because the argument lies in $(-\pi,\pi]$, $\Delta t_{\text{drift}}$ is uniquely recovered as
\begin{equation}
\Delta t_{\text{drift}}
=
\frac{T_{\text{bin}}}{2\pi}\,
\arg\!\left(m_2\overline{m_1}\right).
\end{equation}
Using the relation $\Delta t_{\text{drift}} = t_{\text{drift}}\,T_{\text{int}}$, the clock drift is given by
\begin{align}
t_{\text{drift}}
&=\frac{T_{\text{bin}}}{2\pi T_{\text{int}}}
\arg\!\left(m_2\overline{m_1}\right)\,,
\end{align}
and the use of the principal argument ensures that the recovered drift corresponds
to the smallest signed clock drift between consecutive acquisitions.

\subsection{Removing the time shift}
According to Eq.\,\eqref{TER}, the time shift at the end of the second acquisition is, 
\begin{equation}
    \Delta t(2T_\text{int}) = t_0 + 2t_{\text{drift}}T_\text{int}\,.
\end{equation}
Therefore, the average time of arrival of the third histogram, considering the drift to be perfectly compensated, would be
\begin{equation}
    \mu_e+ \Delta t(2T_\text{int})\,.
\end{equation}
The goal is to add a delay to the TDC such that the time shift $\Delta t(2T_\text{int})$ cancels. The smallest signed delay is
\begin{align}
    \epsilon
    &\equiv \mu_e- \left(\mu_e+\Delta t(2T_\text{int})\bmod {T_\text{bin}}\right) \nonumber\\
    &=\mu_e-\left(\left(\mu_{e}^{[0,T_\text{int})} +\frac{3}{2}\Delta t_{\text{drift}}\right)\bmod{T_{\text{bin}}}\right)\nonumber\\
    &=\mu_e-\left(\left(\mu_{e}^{[T_\text{int},2T_\text{int})}+\frac{1}{2}\Delta t_{\text{drift}}\right)\bmod{T_{\text{bin}}}\right)\,,
\end{align}
where $\Delta t_{\text{drift}}=t_{\text{drift}}T_\text{int}$. Let us now express $\epsilon$ as a function of the circular means. The argument of the circular mean of the first and second histograms can be expressed as
\begin{align}
\begin{cases}
    \arg (m_1)
    =\frac{2\pi \mu_{e}^{[0,T_\text{int})}}{T_\text{bin}} +\phi_q \pmod {2\pi}\,,\\
    \arg (m_2)
    =\frac{2\pi \mu_{e}^{[T_\text{int},2T_\text{int})}}{T_\text{bin}} +\phi_q\pmod {2\pi}\,,
    \end{cases}
\end{align}
where $\phi_q=\arg(C)$. This leads to
\begin{align}
\begin{cases}
\mu_{e}^{[0,T_\text{int})}
     =\frac{T_\text{bin}}{2\pi}(\arg (m_1)-\phi_q) \pmod {T_\text{bin}}\,,\\
\mu_{e}^{[T_\text{int},2T_\text{int})}
    =\frac{T_\text{bin}}{2\pi}(\arg (m_2)-\phi_q) \pmod {T_\text{bin}}\,,    
    \end{cases}
    \label{circ_to_linear}
\end{align}
and 
\begin{align}
    \begin{cases}
\epsilon=\mu_e-\left(\left(\frac{T_{\text{bin}}}{2\pi}\left(\arg(m_1)-\phi_q\right)
+\frac{3}{2}{\Delta t}_{\text{drift}}\right)\bmod{T_{\text{bin}}}\right)\,,\\
\epsilon=\mu_e-\left(\left(\frac{T_{\text{bin}}}{2\pi}\left(\arg(m_2)-\phi_q\right)
+\frac{1}{2}{\Delta t}_{\text{drift}}\right)\bmod{T_{\text{bin}}}\right)\,.\\
    \end{cases}
\end{align}
$\phi_q$ depends on the properties of the SPAD distribution and should be determined independently. Indeed, writing $q(t')$ as the convolution of a Gaussian, a SPAD, and a drift dependent PDF 
\begin{equation}
    q(t')=(\varphi*k*r)(t')\,,
\end{equation}
where $r$ is defined as
\begin{equation}
r(t)=
\begin{cases}
\dfrac{1}{\Delta t_{\text{drift}}}\,\,
& -\Delta t_{\text{drift}}/2 \le t \le \Delta t_{\text{drift}}/2\,,\\[0.8em]
0\,, & \text{otherwise}\,,
\end{cases}
\end{equation}
leads to
\begin{equation}
    \phi_q=\arg\left(\int e^{\frac{2\pi i u}{T_\text{bin}}}k(u)\,du\right)\,,
\end{equation}
since $\varphi$ and $r$ are even, their Fourier coefficients are real and therefore do not contribute to the phase.
Considering the parameters of our SPAD distribution, $\phi_q\approx -0.026$, yielding a bias of approximately $-4.2\,\text{ps}$.

\section{Clock stability requirements}
In this section, we derive the two clock constraints (Eq.\,\eqref{eq:clock_constraint_1} and Eq.\,\eqref{ineq_short_term}).
\subsection{Clock long-term stability}
\label{subsec:calibration}
For a time $t\geq0$, a more general expression of clock drift is given by 
\begin{equation}
    t_{\text{drift}}(t) = t_{\text{drift}}(0) + \frac{dt_{\text{drift}}}{dt} t\,,
\end{equation}
where $t_{\text{drift}}(0)$ is the constant drift that our algorithm targets, and $dt_{\text{drift}}/dt$ represents the clocks aging. Using a time-to-digital converter measurement with the start signal generated by Bob’s clock and the stop generated directly by Alice’s clock, the histogram width over an integration interval $T_\text{int}$ is given by the time shift
\begin{equation}
\begin{aligned}
\Delta t(T_\text{int}) - \Delta t(0)
&= \int_{0}^{T_\text{int}} \frac{d\Delta t}{dt}\, dt \\
&= \int_{0}^{T_\text{int}} t_\text{drift}(t)\, dt \\
&= t_\text{drift}(0)\,T_\text{int}
+ \frac{1}{2}\frac{dt_{\text{drift}}}{dt}\,T_\text{int}^2\,.
\end{aligned}
\end{equation}

For calibration, we can estimate the clock drift by dividing the width of the histogram by the integration time
\begin{align}
    \hat t_{\text{drift}} 
    & = \frac{\Delta t(T_\text{int})-\Delta t(0)}{T_{\text{int}}}
    \label{calibration} \nonumber\\
    & = t_\text{drift}(0) + \frac{1}{2}\frac{dt_{\text{drift}}}{dt} T_\text{int}\,.
\end{align}
The estimated drift is then used to update Alice’s frequency according to
\begin{equation}
    f_A' = f_A(1+\hat t_{\text{drift}})\,.
\end{equation}
After compensation, the residual drift at $t\geq T_\text{int}$ becomes
\begin{align}
   t_{\text{drift,res}}(t) 
   &=\frac{f_B(t)-f_A'}{f_A'} \nonumber\\
    &=\frac{f_A\left(1+t_{\text{drift}}(t)\right) - f_A(1+\hat t_{\text{drift}})}{f_A(1+\hat t_{\text{drift}})} \nonumber\\
   &=\frac{t_{\text{drift}}(t) - \hat t_{\text{drift}}}{1+\hat t_{\text{drift}}}\,.   
\end{align}
For $\hat t_{\text{drift}}\ll1$, this reduces to
\begin{equation}
\begin{aligned}
t_{\text{drift,res}}(t)
&\approx t_{\text{drift}}(t) - \hat t_{\text{drift}} \\
&\approx \frac{dt_{\text{drift}}}{dt}
\left(t - \frac{1}{2}T_{\text{int}}\right),
\qquad t \geq T_{\text{int}}\,,
\end{aligned}
\end{equation}
which is dominated by the aging term. Thus, once calibration is complete, the clock drift comes from aging, and for $t\gg T_\text{int}/2$ is approximately
\begin{equation}
    t_{\text{drift,res}}(t) \approx  \frac{d\,t_{\text{drift}}}{dt} t\,.
\end{equation}

With $t_c$ the elapsed time since calibration, the maximum time allowed from calibration until starting the synchronization algorithm is such that the clock drift--induced time shift over the interval $[t_{c,\text{max}},\,t_{c,\text{max}}+\tau_D]$ remains below $T_\text{bin}/2$, leading to
\begin{equation}
        \left|\frac{d\,t_{\text{drift}}}{dt}\right| \left(t_{c,\text{max}} + \frac{1}{2} \tau_D\right)<\frac{T_\text{bin}}{2\tau_D}\,.
\end{equation}
For $\tau_D\ll 2t_{c,\text{max}}$, we have
\begin{equation}
    \left|t_{\text{drift,res}}\left(t_{c,\text{max}}\right)\right| < \left|t_{\text{drift,max}}\right| \,.
\end{equation}

Finally, one can express the residual clock drift due to aging as a function of the individual aging of Alice's and Bob's clocks. Let the clock drift due to aging since calibration of Alice's and Bob's clocks with respect to a perfect reference be 
\(t_{\text{drift},A}(t_c)\) and \(t_{\text{drift},B}(t_c)\), respectively.  
The drift of Bob's clock relative to Alice's due to aging is then
\begin{equation}
    t_{\text{drift,res}}(t_c)
= \frac{f_B(t_c) - f_A(t_c)}{f_A(t_c)}
= \frac{t_{\text{drift},B}(t_c) - t_{\text{drift},A}(t_c)}
       {1 + t_{\text{drift},A}(t_c)}\,.
\end{equation}
For  \(|t_{\text{drift},A}(t_c)|\ll 1\), this reduces to 
\begin{equation}
    t_{\text{drift,res}}(t_c)
\approx t_{\text{drift},B}(t_c) - t_{\text{drift},A}(t_c)\,.
    \label{drift_res_2_clocks}
\end{equation}
Indeed, we estimated \(|dt_{\text{drift}}/dt|\) of Eq.\,\eqref{eq:clock_constraint_1} and Eq.\,\eqref{ineq_short_term} assuming two identical clocks drifting in opposite directions, so that the relative clock drift rate \(|dt_{\text{drift}}/dt|\) is twice the individual clock drift rate.

\subsection{Clock short-term stability}
\label{subsec:aging_and_synchro}
In the presence of temporal variations of the clock drift, the clock drift--induced time shift after a time t is given by
\begin{equation}
    \Delta t_\text{drift}(t) = \int_0^t t_{\text{drift}}(\tau)\,d\tau
= t_{\text{drift}}(0) t + \frac12 \frac{dt_{\text{drift}}}{dt} t^2\,.
\end{equation}
Through the mean arrival time of the photons, our algorithm calculates the mean clock drift--induced time shift over the first and second integration windows,
$[0,T_{\text{int}})$ and $[T_{\text{int}},2T_{\text{int}})$, which are
\begin{align}
m_1 = \frac{1}{T_{\text{int}}}\int_0^{T_{\text{int}}} \Delta t_\text{drift}(t)\,dt\,,\\
m_2 = \frac{1}{T_{\text{int}}}\int_{T_{\text{int}}}^{2T_{\text{int}}} \Delta t_\text{drift}(t)\,dt\,. 
\end{align}
A straightforward calculation yields
\begin{equation}
    m_2 - m_1 = t_{\text{drift}}(0)T_{\text{int}} + \frac{dt_{\text{drift}}}{dt}T_{\text{int}}^2 \,,
\end{equation}
and the clock drift estimator of the synchronization algorithm in the presence of aging becomes
\begin{equation}
    \hat t_{\text{drift}}=\frac{m_2 - m_1}{T_{\text{int}}}
= t_{\text{drift}}(0) + \frac{dt_{\text{drift}}}{dt} T_{\text{int}}\,,
\end{equation}
meaning that the estimator picks up both the constant drift $t_{\text{drift}}(0)$ and a contribution proportional to the time derivative. After frequency compensation, the residual drift becomes
\begin{equation}
    t_{\text{drift,res}}(t)
\approx t_{\text{drift}}(t) - \hat t_{\text{drift}}
\approx \frac{dt_{\text{drift}}}{dt} (t - T_{\text{int}})\,.
\end{equation}

For the next two histograms to remain below a chosen error threshold $\epsilon^\text{(w)}_{\text{thr}}$, we need the residual clock drift-induced time shift over $[T_{\text{int}}',2T_{\text{int}}')$, with $T_\text{int}'\equiv2T_\text{int}$, to be below $\Delta t_\text{drift}(\epsilon^\text{(w)}_{\text{thr}}).$ The residual clock drift--induced time shift over $[T_{\text{int}}',2T_{\text{int}}') $ is
\begin{align}
\Delta t_{\text{res}}(2T_{\text{int}}')
&= \int_{T_{\text{int}}'}^{2T_{\text{int}}'}
    t_{\text{drift,res}}(t)\,dt \nonumber\\
&= \frac{d t_{\text{drift}}}{dt}
    \left( \frac{3T_{\text{int}}'^2}{2} - T_{\text{int}}T_{\text{int}}'\right)\,.
\end{align}
Using $T_{\text{int}}' = 2T_{\text{int}}$, we obtain
\begin{equation}
\Delta t_{\text{res}}(2T_{\text{int}}')
= 4T_{\text{int}}^2\,\frac{d t_{\text{drift}}}{dt}\,.
\end{equation}
Thus, a constraint on the short-term clock stability is
\begin{equation}
    4T_{\text{int}}^2\,
\left|\frac{d t_{\text{drift}}}{dt}\right|
\;<\;
\Delta t_\text{drift}(\epsilon^{\text{(w)}}_{\text{thr}})\,.
\end{equation}
Note that in principle, $\Delta t_{\text{res}}$ remains the same for the next iterations, provided that we follow the drift estimation and compensation procedure.

\section{Effect of Poissonian noise on clock-drift estimation}
\label{subsec:Relative_error_drift}

To evaluate the effect of photon counting noise on the clock-drift estimation, we generate histograms folded over $T_\text{hist}=T_\text{bin}=1\,\text{ns}$ from the analytical PDF of the photon detection time (Eq.\,\eqref{eq:ptotspad}). As explained in Sec.\ref{subsec:pdf_folded_hist_during_sync}, the PDF reduces to $\tilde p_\text{e,SPAD}(\tau)$, where $\tau = t' \bmod{T_\text{bin}}$.

Defining the histogram binning with a bin width $\Delta \tau=100\,\text{ps}$, the probability to detect a photon in the $i$-th bin is obtained by integrating the PDF over the corresponding time interval
\begin{equation}
P_i = \int_{\tau_i}^{\tau_i+\Delta \tau} \tilde p_\text{e,SPAD}(\tau)\, d\tau\,.
\end{equation}
For a given integration time $T_{\mathrm{int}}$, the expected number of signal counts in bin $i$ is
\begin{equation}
\lambda_i^{\mathrm{sig}} = T_{\mathrm{int}} \, \text{cps}_{\mathrm{eff,Alice}} \, P_i\,,
\end{equation}
where $\text{cps}_{\mathrm{eff,Alice}}$ is the signal photon detection rate (Eq.\,\eqref{eq:cps_eff_Alice}). Dark counts are modeled as a uniform background over the histogram bins, yielding an additional contribution
\begin{equation}
\lambda_i^{\mathrm{dc}} = T_{\mathrm{int}} \, \frac{\text{cps}_{\mathrm{eff,dc}}}{N_{\mathrm{bins}}}\,,
\end{equation}
where $\text{cps}_{\mathrm{eff,dc}}$ is the effective dark count rate (Eq.\,\eqref{eq:cps_eff_DC}) and $N_{\mathrm{bins}}$ is the total number of histogram bins.
The expected number of counts in bin $i$ is therefore
\begin{equation}
\lambda_i = T_{\mathrm{int}} \left( \text{cps}_{\mathrm{eff,Alice}} P_i + \frac{\text{cps}_{\mathrm{eff,dc}}}{N_{\mathrm{bins}}} \right)\,,
\end{equation}
and a histogram is obtained by drawing the number of counts in each bin from a Poisson distribution with mean $\lambda_i$,
\begin{equation}
H_i \sim \mathrm{Poisson}(\lambda_i)\,.
\end{equation}

\begin{figure}[htbp]
\centering

\begin{subfigure}{\linewidth}
    \centering
    \includegraphics[width=\linewidth]{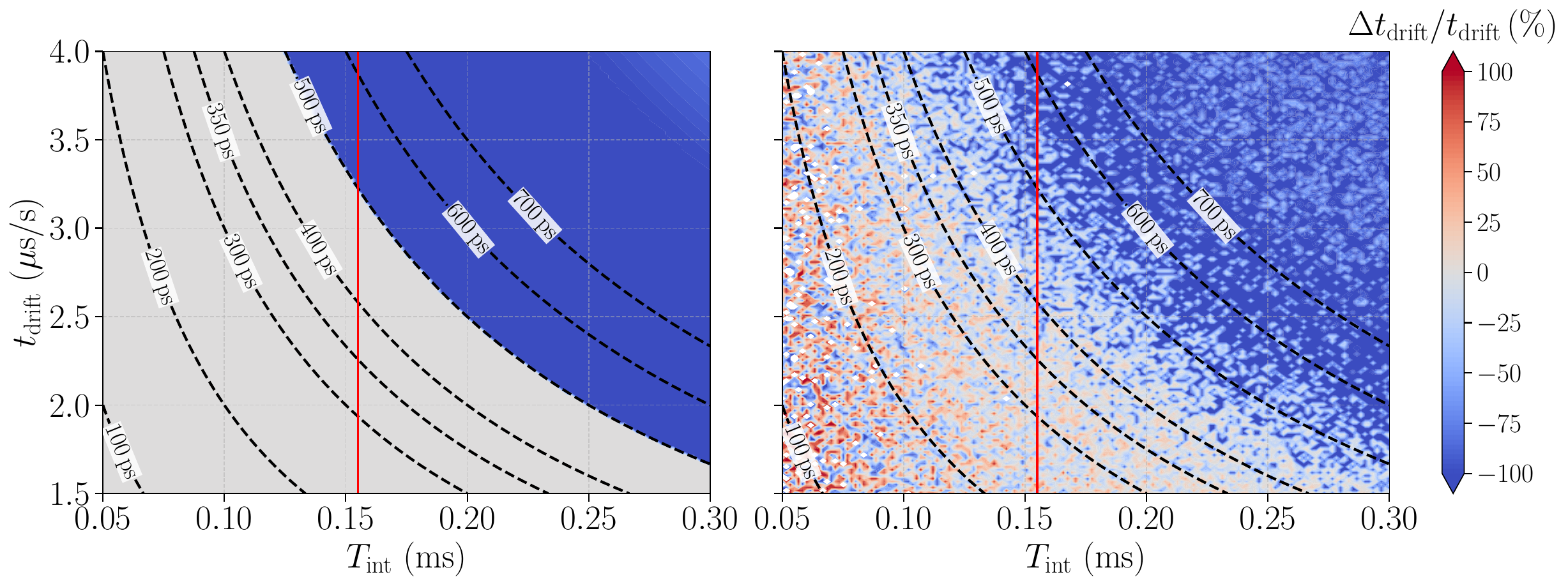}
    \caption{Early synchronization stage ($\bar n=10$, short integration times).}
\end{subfigure}
\vspace{6pt}
\begin{subfigure}{\linewidth}
    \centering
    \includegraphics[width=\linewidth]{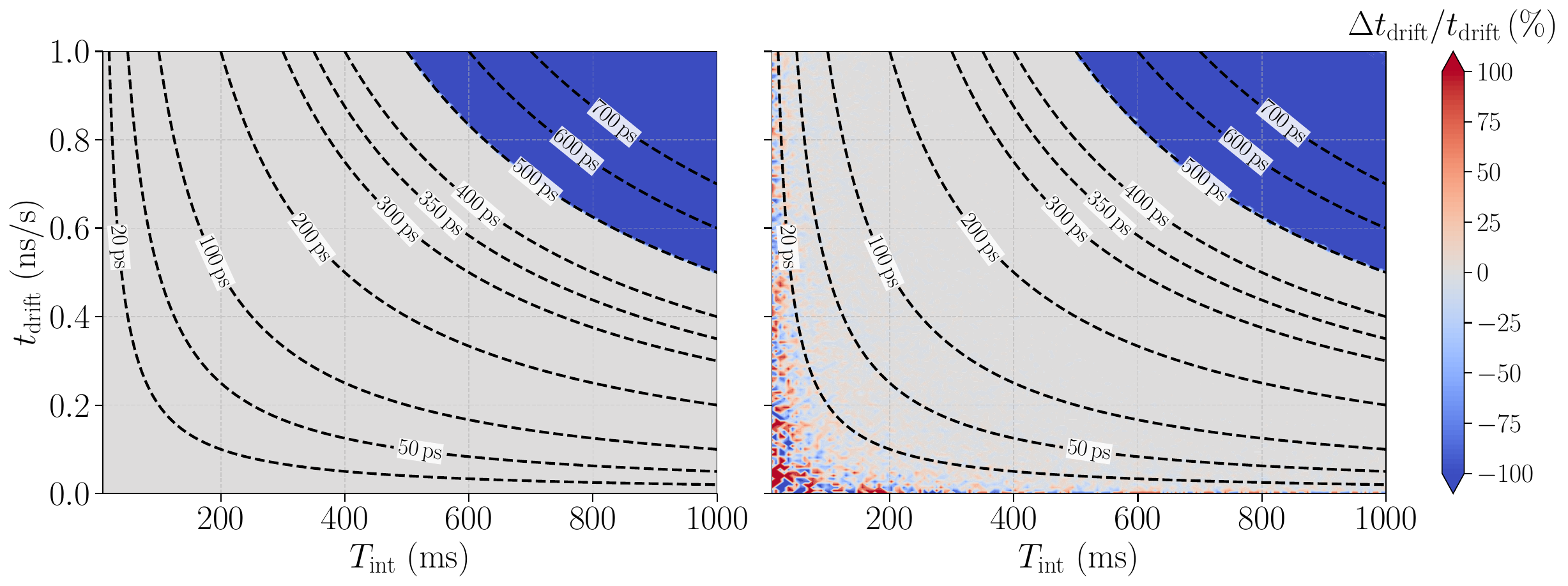}
    \caption{Final synchronization stage ($\bar n=0.225$).}
\end{subfigure}

\caption{Relative error in clock-drift estimation as a function of the true drift $t_{\mathrm{drift}}$ (vertical axis) and the histogram integration time $T_{\mathrm{int}}$ (horizontal axis). In each panel, the left subfigure shows the noiseless simulation, while the right subfigure includes Poissonian counting noise. The color scale represents the relative estimation error $(\hat t_{\mathrm{drift}}-t_{\mathrm{drift}})/t_{\mathrm{drift}}$ in percent. Dashed curves correspond to isolines of $\Delta t_{\mathrm{drift}} = t_{\mathrm{drift}} T_{\mathrm{int}}$. The vertical line indicates the integration time $T_{\mathrm{int}}=155~\mu\mathrm{s}$ used at the beginning of the synchronization algorithm.}

\label{fig:simu_drift_all}
\end{figure}

Figure~\ref{fig:simu_drift_all} shows the relative error between the estimated clock drift $\hat{t}_{\mathrm{drift}}$ and the true clock drift $t_{\mathrm{drift}}$ for a mean photon number $\bar n=10$ and at a distance $z=120\,\text{km}$.
The estimate $\hat{t}_{\mathrm{drift}}$ is obtained by applying the synchronization algorithm to two simulated histograms generated for a given $t_{\mathrm{drift}}$.  In the absence of Poissonian noise, the clock drift is recovered exactly over the entire parameter space. This accuracy stems from the broadness of the photon detection-time PDF compared to the histogram bin width of $100$ ps. 

When Poissonian noise is included, statistical fluctuations in the photon counts blur the sharp transition at $\Delta t_\text{drift}(T_\text{int})=T_{\mathrm{bin}}/2 = 500\,\text{ps}$. In practice, $\hat t_\text{drift}$ may cross this boundary of ambiguity prematurely, leading to incorrect drift estimates. To ensure robust operation, we restrict the maximum clock drift--induced time shift to $70\,\%$ of this limit, corresponding to approximately $\Delta t_\text{drift}(T_\text{int})=350\,\text{ps}$.

For completeness, we simulate the impact of Poissonian noise for the final stage of the algorithm, where the integration time reaches $T_{\mathrm{int}} = 0.5\,\text{s}$ and the mean photon number is $\bar n = 0.225$. In this regime, the signal-to-noise ratio is significantly improved, and the algorithm can reliably recover clock drifts up to $\pm40\,\text{ps}/\text{s}$ at $T_\text{int}=0.5\,\text{s}$ with only a small relative error.

\section{Initializing the clock drift}
\label{subsec:init_clock_drift}
In experiments, Alice’s clock frequency is set so that the initial clock drift is $\left|t_{\text{drift,max}}^{(\text{practical})}\right|$. Defining the drift before this calibration as $t_{\text{drift}}=(f_B-f_A)/f_A$ and after calibration as $\left|t_{\text{drift,max}}^{(\text{practical})}\right|=(f_B-f_A')/f_A'$, Alice’s calibration frequency is
\begin{equation}
    f_A' = f_A \frac{1 + \hat t_{\text{drift}}}{1 + \left|t_{\text{drift,max}}^{(\text{practical})}\right|}\,,
\end{equation}
where $\hat t_{\text{drift}}$ experimentally estimates $t_{\text{drift}}$ (Eq.~\eqref{calibration}).

\section{Convergence of the synchronization algorithm}
\label{subsec:sync_fast}
\begin{figure}[htbp]
  \centering
  \includegraphics[width=\linewidth]{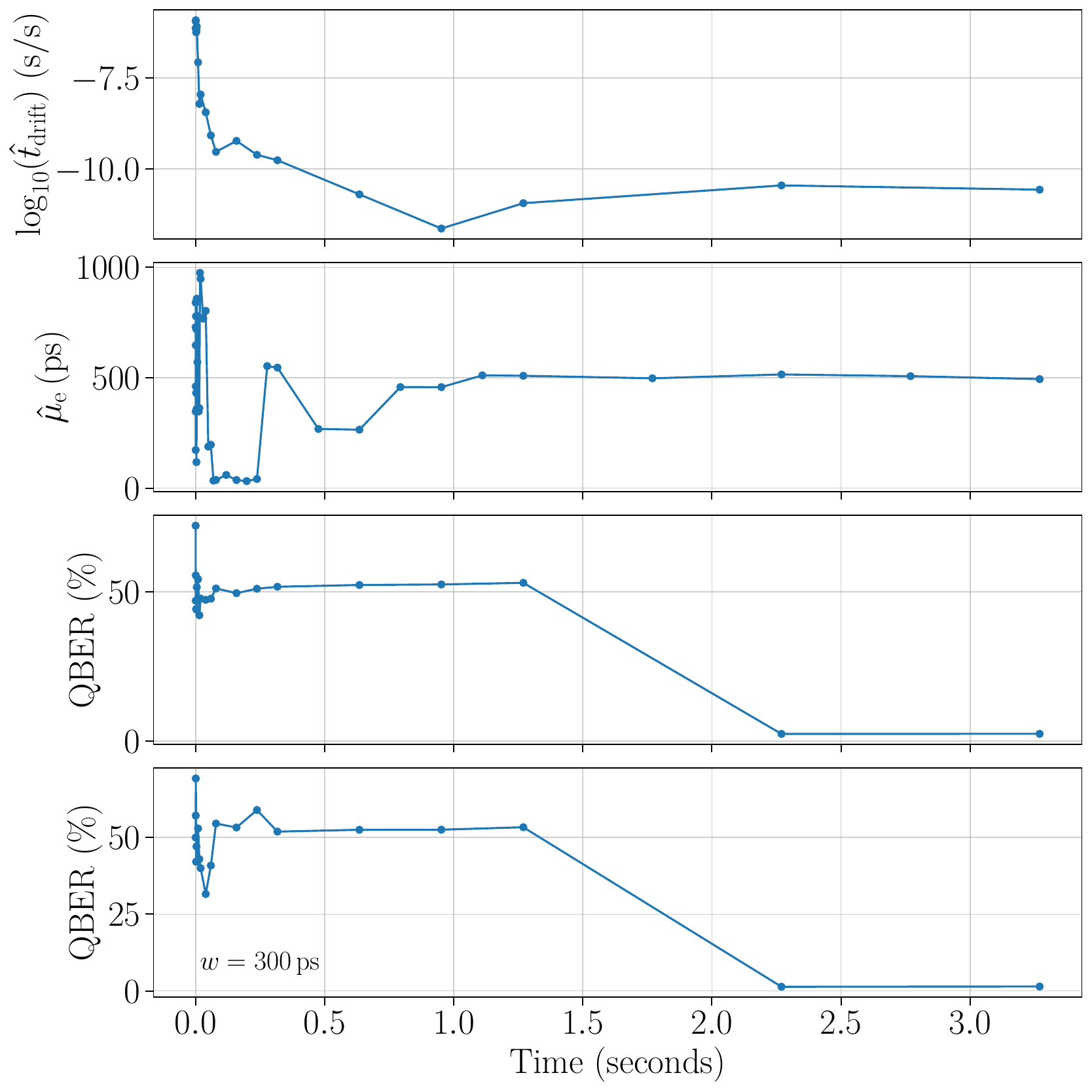}
  \caption{Clock drift $\hat t_\text{drift}$, mean photon detection time $\hat \mu_\text{e}$, QBER, and filtered QBER with $w=300\,\text{ps}$ at the beginning of the synchronization.}
  \label{fig:sync_fast}
\end{figure}

Figure~\ref{fig:sync_fast} shows the initial stage of synchronization as a function of the cumulative histogram acquisition time. The clock drift is initialized at $\left|t_{\text{drift,max}}^{(\text{practical})}\right|\approx 2.3\,\mu\text{s}/\text{s}$. Histograms are initially acquired over $T_{\text{int}}=155\,\mu\text{s}$, and $T_{\text{int}}$ is quadrupled after every three iterations until $500\,\text{ms}$. Clock offset recovery is performed right before $T_\text{int}$ reaches 500\,ms, which corresponds to approximately $1.3\,\text{s}$ in cumulative acquisition time. Starting from the next pair of histograms, the QBER drops from $50\,\%$ to its nominal value. Note that the cumulative acquisition time could be reduced by increasing $T_{\text{int}}$ more aggressively. However, for a given initial drift, excessive increase in $T_{\text{int}}$, combined with clock-drift estimation errors due to Poissonian statistics, can flatten the histogram and impair drift estimation.

\clearpage
\bibliography{IF_BB84}     
\end{document}